\documentclass[12pt,draftcls,onecolumn]{IEEEtran}

\renewcommand{\P}{\mbox{P}}
\renewcommand{\P}{{\mathbb{P}}}

\newcommand{\beq}{\begin{equation}}
\newcommand{\eeq}{\end{equation}}
\newcommand{\beqa}{\begin{eqnarray}}
\newcommand{\eeqa}{\end{eqnarray}}
\newcommand{\dfz}{\triangleq}

\newcommand{\bd}{\bm{d}}
\newcommand{\bx}{\bm{x}}
\newcommand{\by}{\bm{y}}
\newcommand{\bz}{\bm{z}}
\newcommand{\bv}{\bm{v}}

\newcommand{\VAR}{\textnormal{VAR}}

\usepackage{cite}
\usepackage{amsmath} 
\usepackage{amssymb} 
\usepackage{dsfont}

\usepackage{graphicx}
\usepackage{graphics}
\usepackage{bm}
\input{epsf}

\usepackage{color}

\newcommand{\E}{{\mathbb{E}}}

\begin{document}
%
\title{Diffusion-Based Adaptive Distributed Detection: Steady-State Performance in the Slow Adaptation Regime}

\author{Vincenzo~Matta, Paolo~Braca, Stefano~Marano, Ali~H.~Sayed
\thanks{
The work of A.~H.~Sayed  was supported in part by NSF grants CCF-1011918, CCF-1524250, and ECCS-1407712. 
A short and limited version of this work appears in the conference publication~\cite{ADD_ICASSP2014}.
}
\thanks{
V.~Matta and S.~Marano are with DIEM, University of Salerno, via Giovanni Paolo II 132, I-84084, Fisciano (SA), Italy (e-mail: vmatta@unisa.it; marano@unisa.it).

P.~Braca is with NATO STO Centre for Maritime Research and Experimentation, La Spezia, Italy (e-mail: braca@cmre.nato.int).

A.~H.~Sayed is with the Electrical Engineering Department, University of California, Los Angeles, CA 90095 USA (e-mail: sayed@ee.ucla.edu).
}
}

\maketitle

\begin{abstract}
This work examines the close interplay between cooperation and adaptation for distributed detection schemes over fully decentralized networks. The combined attributes of cooperation and adaptation are necessary to enable networks of detectors to continually learn from streaming data and to continually track drifts in the state of nature when deciding in favor of one hypothesis or another. The results in the paper establish a fundamental scaling law for the steady-state probabilities of miss-detection and false-alarm in the slow adaptation regime, when the agents interact with each other according to distributed strategies that employ small constant step-sizes. The latter are critical to enable continuous adaptation and learning. The work establishes three key results. First, it is shown that the output of the collaborative process at each agent has a steady-state distribution.  Second, it is shown that this distribution is asymptotically Gaussian in the slow adaptation regime of small step-sizes. And third, by carrying out a detailed large deviations analysis, closed-form expressions are derived for the decaying rates of the false-alarm and miss-detection probabilities. Interesting insights are gained from these expressions. In particular, it is verified that as the step-size $\mu$ decreases, the error probabilities are driven to zero exponentially fast as functions of $1/\mu$, and that the exponents governing the decay increase linearly in the number of agents. It is also verified that the scaling laws governing errors of detection and errors of estimation over networks behave very differently, with the former having an exponential decay proportional to $1/\mu$, while the latter scales linearly with decay proportional to $\mu$. Moreover, and interestingly, it is shown that the cooperative strategy allows each agent to reach the same detection performance, in terms of detection error exponents, of a centralized stochastic-gradient solution. The results of the paper are illustrated  by applying them to canonical distributed detection problems.
\end{abstract}


\section{Overview}
\IEEEPARstart{R}{ecent} advances in the field of distributed inference have produced several useful strategies aimed at exploiting local {\em cooperation} among network nodes to enhance the performance of each individual agent. 
However, the increasing availability of streaming data continuously flowing across the network has added the new and challenging requirement of online {\em adaptation} to track drifts in the data. In the adaptive mode of operation, the network agents must be able to enhance their learning abilities continually in order to  produce reliable inference in the presence of drifting  statistical conditions, drifting  environmental conditions, and even changes in the network topology, among other possibilities.  Therefore, concurrent adaptation (i.e., tracking) and learning (i.e., inference) are key components for the successful operation of distributed networks tasked to produce reliable inference under dynamically varying conditions and in response to streaming data.  

Several useful distributed implementations based on consensus strategies~\cite{running-cons,asymptotic-rc,Bracaetal-Pageconsensus,MouraLDGauss,MouraLDnonGauss,MouraLDnoisy,TsitsiklisBertsekasAthans,xiao-boyd,BoydGhoshetal,Nedic,Dimakis,KarMoura} and diffusion strategies~\cite{LopesSayed,CattivelliSayedEstimation,SayedSPmag,SayedAcademicPress,SayedProcIEEE,ChenSayed} have been developed for this purpose in the literature. The diffusion strategies have been shown to have superior stability ranges and mean-square performance when constant step-sizes are used to enable continuous adaptation and learning~\cite{TuSayedConsensus}. For example, while consensus strategies can lead to unstable growth in the state of adaptive networks even when all agents are individually stable, this behavior does not  occur for diffusion strategies. In addition, diffusion schemes are robust, scalable, and fully decentralized. Since in this work we focus on studying {\em adaptive} distributed inference strategies, we shall therefore focus on diffusion schemes due to their enhanced mean-square stability properties over adaptive networks.

Now, the  interplay between the two fundamental aspects of cooperation and adaptation has been investigated rather extensively in the context of {\em estimation} problems. Less explored in the literature is the same interplay in the context of {\em detection} problems. This is the main theme of the present work. Specifically, we shall address the problem of designing and characterizing the performance of diffusion strategies that reconcile both needs of adaptation and detection in decentralized systems. The following is a brief  description of the scenario of interest. 

\begin{figure}[t]
\centerline{\includegraphics[width=.55\textheight]{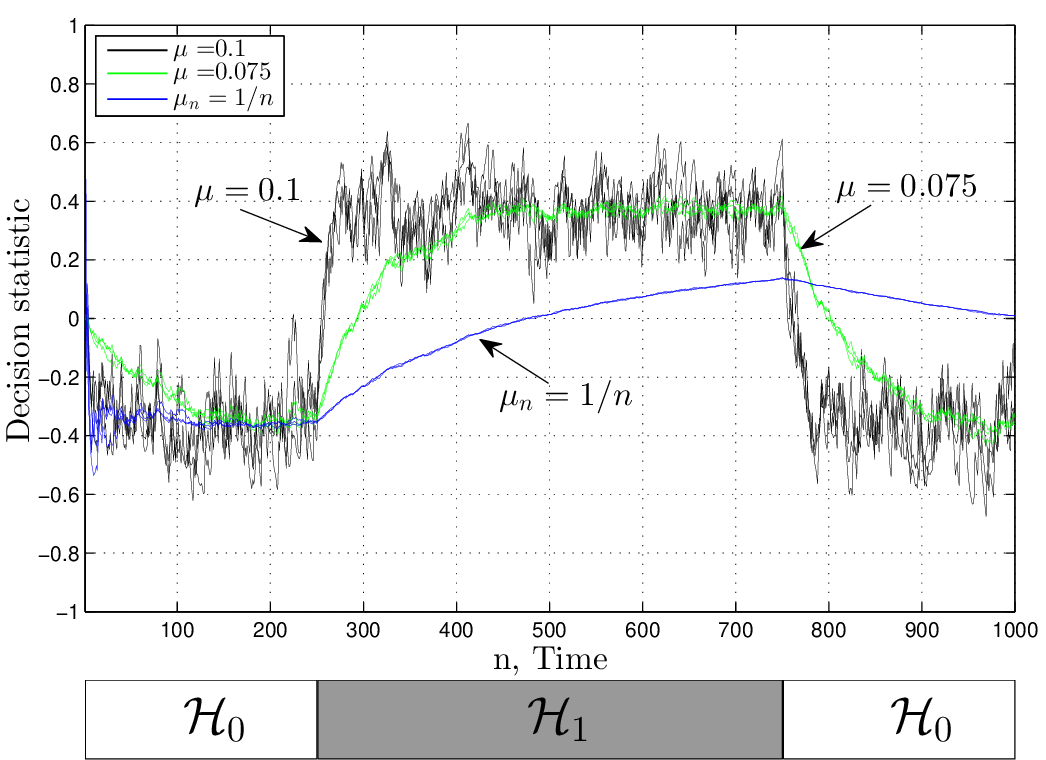}}
\caption{The top panel illustrates the time-evolution of the decision statistics at three generic local agents for two situations: (a) constant step-size adaptation using a diffusion strategy  and (b) diminishing step-size updates using $\mu_n=1/n$ and a running consensus strategy. The actual variation of the true hypothesis is depicted in the bottom panel from ${\cal H}_0$ to ${\cal H}_1$ to ${\cal H}_0$.}
\label{fig:adaptection}
\end{figure}

A network of connected agents is assumed to monitor a certain phenomenon of interest. As time elapses, the agents collect an increasing amount of streaming data, whose statistical properties depend upon an {\em unknown} state of nature. The state is formally represented by a pair of hypotheses, say, ${\cal H}_0$ and ${\cal H}_1$.
At each time instant, each agent is expected to produce a decision about the state of nature, based upon its own observations and the exchange of information with neighboring agents.
The emphasis here is on {\em adaptation}: we allow the true hypothesis to drift over time, and the network must be able to track the drifting state. 
This framework is illustrated in Fig.~\ref{fig:adaptection}, where we show the time-evolution of the actual realization of the decision statistics computed by three generic network agents. Two situations are considered. In the first case, the agents run a constant-step size diffusion strategy~\cite{ZhaoSayedLMSestimation,SayedSPmag} and in the second case, the agents run a consensus strategy with a diminishing step-size of the form $\mu_n=1/n$~\cite{running-cons,asymptotic-rc,Bracaetal-Pageconsensus,MouraLDGauss,MouraLDnonGauss,MouraLDnoisy}. Note from the curves in the figure that the statistics computed by different sensors are hardly distinguishable, emphasizing a certain equivalence in performance among distinct agents, an important feature that will be extensively commented on in the forthcoming analysis.

Assume that high (positive) values of the statistic correspond to deciding for  ${\cal H}_1$, while low (negative) values correspond to deciding for ${\cal H}_0$.
The bottom panel in the figure shows how the true (unknown) hypothesis changes at certain (unknown) epochs following the sequence ${\cal H}_0\rightarrow{\cal H}_1\rightarrow{\cal H}_0$. 
It is seen in the figure that the adaptive diffusion strategy is more apt in tracking the drifting state of nature. It is also seen that the diminishing step-size consensus implementation is unable to track the changing conditions. Moreover, the inability to track the drift degrades further as time progresses since the step-size sequence $\mu_n=1/n$ decays to zero as $n\rightarrow\infty$. For this reason, in this work we shall set the step-sizes to constant values to enable continuous adaptation and learning by the distributed network of detectors. 
In order to evaluate how well these adaptive networks perform, we need to be able to assess the goodness of the inference performance (reliability of the decisions), so as to exploit the trade-off between adaptation and learning capabilities. This will be the main focus of the paper.

\subsection{Related Work}

The literature on distributed detection is definitely rich, see, e.g.,~\cite{LongoLookabaughGray,VarshneyBook,ViswanathanVarshney,BlumKassamPoor,ChamberlandVeeravalli,ChamberlandVeeravalliSPMag,ChenTongVarshneySPMag,Saligrama} as useful entry points on the topic. 
A distinguishing feature of our approach is its emphasis on {\em adaptive} distributed detection techniques that respond to streaming data in real-time. We address this challenging problem with reference to the {\em fully decentralized} setting, where no fusion center is admitted, and the agents cooperate through local interaction and consultation steps. 

For several useful formulations of distributed point estimation and detection, the use of stochastic approximation consensus-based solutions with {\em diminishing} step-sizes leads to asymptotically optimal performance, either in the sense of asymptotic variance in point estimation~\cite{KarMoura}, in the sense of error exponents~\cite{MouraLDGauss,MouraLDnonGauss,MouraLDnoisy}, or in the sense of asymptotic relative efficiency in the locally optimum detection framework~\cite{asymptotic-rc}. Optimality in these works is formulated in reference to the centralized solution, and the qualification ``asymptotic'' is used to refer either to a large number of observations or a large time window. The error performance (e.g., mean-square error for estimation or error probabilities for detection) is shown in these works to decay with optimal rates as time elapses, provided that some conditions on the network structure are met. 
For these results to hold, it is critical for the statistical properties of the data to remain invariant and for the algorithms to rely on a recursive test statistics with a {\em diminishing} step-size.

In some other distributed inference applications, however, the statistical properties of the data can vary over time. For instance, in a detection problem, the actual hypothesis in force, and/or some parameters of the pertinent distributions, might change at  certain moments. Therefore, the adaptation aspect, i.e., the capability of persistently tracking dynamic scenarios, becomes important. In such scenarios, the diffusion algorithms (with non-diminshing, constant step-size) provide effective mechanisms for continuous adaptation and learning. Similar to the consensus-based algorithms with diminishing step-sizes, they are easy to implement, since they involve linear operations, and are naturally suited to a fully distributed implementation. However, differently from the consensus algorithms with diminishing step-size, the strategies with constant step-size are inherently able to work under dynamically changing conditions and offer enhanced tracking capability.

\subsection{Inherent Tracking Mechanism}
It is well-known in the adaptation and learning literature that using {\em constant step-sizes} in the update relations automatically infuses the algorithms with a tracking mechanism that enables them to track variations in the underlying models. This is because constant step-sizes keep adaptation alive, forever. This is in contrast to decaying step-sizes, which tend to zero and ultimately stop adapting. With a constant step-size, learning is always active. When the hypothesis changes, an algorithm with a constant step-size will continue learning from that point onwards and given sufficient time to learn, the steady-state analysis in this article will show that the probabilities of error will indeed decay exponentially as functions of the inverse of the step-size.  

The key challenge in these scenarios is that a constant step-size keeps the update active, which then causes gradient noise to seep continuously into the operation of the algorithm. This effect does not happen for decaying step-sizes because the diminishing step-size annihilates the gradient noise term in the limit. However, a decaying step-size cannot track changing hypotheses due to the vanishing step-size. The difficulty in the constant step-size case is therefore to show that despite the presence of gradient noise, the dynamics of the learning algorithm is such that it can keep this effect under check and is capable to learn. The more it learns, the more it reduces the size of the gradient noise and this feedback mechanism leads to effective learning.  This is one of the key conclusions in this work, namely, showing that indeed the probabilities of error decay exponentially with the inverse of the step-size. This result is non-trivial and the derivations will take some effort before arriving at the insightful scaling laws that we are presenting in this work.

\subsection{Analysis of Detection Performance}
The aforementioned properties of the diffusion strategies used in this work explain their widespread utilization in the context of adaptive estimation~\cite{SayedProcIEEE}, and motivate their use in the context of adaptive distributed detection~\cite{CattivelliSayedDetection, ADD_ICASSP2014, ADD_ICASSP2015}. With reference to this class of algorithms, while several results have been obtained for the mean-square-error (MSE) {\em estimation} performance of adaptive networks~\cite{ZhaoSayedLMSestimation,SayedSPmag}, less is known about the performance of distributed detection networks. In particular, in~\cite{CattivelliSayedDetection}, the miss-detection and false-alarm probabilities have been evaluated with reference to Gaussian observations. 
However, a detailed analytical characterization of the detection performance (i.e., false-alarm and detection probabilities), with reference to a general observational model, is still missing.
This is mainly due to the fact that results on the asymptotic distribution of the error quantities under constant step-size adaptation over networks are largely unavailable in the literature. 

While reference~\cite{LMSdist} argues that the error  in single-agent least-mean-squares (LMS) adaptation converges in distribution, the resulting distribution is not characterized. These questions are considered in~\cite{ZhaoSayedSSP2011, ChenSayedGlobalSIP} in the context of distributed estimation over adaptive networks. Nevertheless, these results on the asymptotic distribution of the errors are still insufficient to characterize the rate of decay of the probability of error over networks of distributed detectors. The main purpose of this work is to fill this gap.
To do so, it is necessary to pursue a large deviations analysis in the constant step-size regime. Motivated by these remarks, we therefore provide a thorough statistical characterization of the diffusion network in a manner that enables detector design and analysis.

\vspace*{10pt}
\noindent
{\bf Notation.} We use boldface letters to denote random variables, and normal font letters for their realizations. Capital letters refer to matrices, small letters to both vectors and scalars. 
Sometimes we violate this latter convention, for instance, we denote the total number of sensors by $S$. The symbols $\P$ and $\E$ are used to denote the probability and expectation operators, respectively. 
The notation $\P_h$ and $\E_h$, with $h=0,1$, means that the pertinent statistical distribution corresponds to hypothesis ${\cal H}_0$ or ${\cal H}_1$.

\section{Preliminaries and Main Results}
\label{sec:probform}
Consider a connected network of $S$ agents.
The scalar observation collected by the $k$-th sensor at time $n$ will be denoted by $\bx_k(n)$, $k=1,2,\dots,S$. 
Data are assumed to be spatially and temporally independent and identically distributed (i.i.d.), {\em conditioned} on the hypothesis that gives rise to them. The distributed network is interested in making an inference about the true state of nature (i.e., the underlying hypothesis), which is allowed to vary over time. Since in this work we focus on a steady-state analysis, it is unnecessary at this stage to introduce an explicit dependence of the datum $\bx_k(n)$ on the particular hypothesis giving rise to it.
\\

\noindent 
{\bf Remark.}
When dealing with i.i.d. observations across sensors, the important issue of local versus aggregate distinguishability is bypassed. In most practical scenarios, sensors observe different aspects of a field, so local distinguishability is hard to achieve but the collective observation model may still be globally informative. The issue when local information is not sufficient for discrimination has been studied in several works before, including~\cite{SpatialRef1,SpatialRef2,SpatialRef3}, and in other related references on diffusion strategies. In the context of multi-agent processing, the distinguishability condition essentially amounts to a positivity condition on the global Gramian (Hessian) matrix while allowing the individual Gramians to be non-negative definite. Learning is still possible in these cases, as shown, for example, in~\cite{ChenSayedIT2015Part1,ChenSayedIT2015Part2,SayedProcIEEE}. 
~\hfill$\square$
\vspace*{3pt}

As it is well-known, for the i.i.d. data model, an optimal centralized (and non-adaptive) detection statistic is the sum of the log-likelihoods. When these are not available, alternative detection statistics obtained as the sum of some suitably chosen functions of the observations are often employed, as happens in some specific frameworks, e.g., in locally optimum detection~\cite{kassam} and in universal hypothesis testing~\cite{poorbook}. Accordingly, each sensor in the network will try to compute, as its own detection statistic, a weighted combination of some function of the local observations. 
We assume the symbol $\bx_k(n)$ represents the local statistic that is available at time $n$ at sensor $k$.

Since we are interested in an adaptive inferential scheme, and given the idea of relying on weighted averages, we resort to the class of diffusion strategies for adaptation over networks~\cite{CattivelliSayedDetection,SayedSPmag}. These strategies admit various forms. We consider the ATC form due to some inherent advantages in terms of a slightly improved mean-square-error performance relative to other forms~\cite{SayedSPmag}. In the ATC diffusion implementation, each node $k$ updates its state from $\by_{k}(n-1)$ to $\by_{k}(n)$ through local cooperation with its neighbors as follows:
\begin{eqnarray}
\bv_k(n)&=&\by_k(n-1) + \mu[\bx_k(n)-\by_k(n-1)],\label{eq:diff1}\\
\by_k(n)&=&\sum_{\ell=1}^S a_{k,\ell} \bv_{\ell}(n)
\label{eq:diff2}
\end{eqnarray}
where $0<\mu\ll 1$ is a small step-size parameter. In this construction, node $k$ first uses its local statistic, $\bx_k(n)$, to update its state from $\by_{k}(n-1)$ to an intermediate value $\bv_k(n)$. All other nodes in the network perform similar updates simultaneously using their local statistics. Subsequently, node $k$ aggregates the intermediate states of its neighbors using nonnegative convex combination weights $\{a_{k,\ell}\}$ that add up to one. Again, all other nodes in the network perform a similar calculation. If we collect the combination coefficients into a matrix~$A=[a_{k,\ell}]$, then $A$ is a right-stochastic matrix in that the entries on each of its rows add up to one:
\beq
a_{k,\ell}\geq 0,\quad A\mathds{1}=\mathds{1},
\eeq
with $\mathds{1}$ being a column-vector with all entries equal to $1$.

\subsection{Performance and Convergence Analyses}
At time $n$, the $k$-th sensor needs to produce a decision based upon its state value $\by_k(n)$.
To this aim, a decision rule must be designed, by choosing appropriate decision regions.
The performance of the test will be measured according to the Type-I (false-alarm) and Type-II (miss-detection) error probabilities defined, respectively, as
\beqa
\alpha_k(n)&\dfz&\P\left[
\begin{array}{c}
\textnormal{agent $k$ decides ${\cal H}_1$ at time $n$ }\\
\textnormal{while ${\cal H}_0$ is true at time $n$}
\end{array}
\right],\label{eq:alphaknfirstdef}\\
\beta_k(n)&\dfz&\P\left[
\begin{array}{c}
\textnormal{agent $k$ decides ${\cal H}_0$ at time $n$ }\\
\textnormal{while ${\cal H}_1$ is true at time $n$}
\end{array}
\right].\label{eq:betaknfirstdef}
\eeqa
Note that these probabilities depend upon the statistical properties of the {\em whole} set of data used in the diffusion algorithm up to current time $n$. In particular, the error probabilities depend upon the different variations of the statistical distributions may have occurred during the evolution of the algorithm, and not only upon the particular hypothesis in force at time $n$. 

Therefore, a rigorous analytical characterization of the system in terms of its overall inference performance at each time instant, and under general operation modalities (i.e., for arbitrarily varying statistical conditions) is generally not viable. This implies, among other difficulties, that the structure of the optimal, or even a reasonable test, is unknown. A standard approach in the adaptation literature to get useful performance metrics and meaningful insights, consists of splitting the analysis in two parts: 
\begin{itemize}
\item[$i)$] A {\em transient} analysis where, starting from a given state, some variations in the statistical conditions occur and the time to track such variations is evaluated. It is possible to carry out studies that focus on the transient phase of the learning algorithm, and to clarify its behavior during this stage of operation, as is done in~\cite{ChenSayedIT2015Part1,ChenSayedIT2015Part2}. 
\item[$ii)$] A {\em steady-state} analysis, where the inference performance is evaluated with reference to an infinitely long period of stationarity. Even in the steady-state regime, an exact analytical characterization of the inference performance is seldom affordable. Therefore, closed-form results are usually obtained working in the regime of slow adaptation, i.e., of small step-sizes. 
\end{itemize}
These two views are complementary. Typically, for a given value of the step-size $\mu$, the diffusion algorithm exhibits the following features: 
\begin{itemize}
\item[$i)$] The convergence rate towards the steady-state regime is known to occur at an exponential rate in the order of $O(c^n)$ for some $c\in (0,1)$; this is a faster rate than $O(1/n)$ that is afforded, for example, by diminishing step-sizes. Nevertheless, in the constant step-size case, the smaller the value of $\mu$ is, the closer the value of $c$ gets to one.  
\item[$ii)$] The steady-state inference performance is a decreasing function of the step-size. Therefore, the lower $\mu$ is, the lower the steady-state error. 
\end{itemize}
In this article, we address in some detail the steady-state performance of diffusion strategies for distributed detection over adaptive networks. Our main interest is in showing that the multi-agent network is able to learn well, with error probabilities exhibiting an exponential decay as functions of $1/\mu$. In particular, our analysis will be conducted with reference to the steady-state properties (as $n\rightarrow\infty$), and for small values of the step-size ($\mu\rightarrow 0$).
Throughout the paper, the term steady-state will refer to the limit as the time-index $n$ goes to infinity, while the term asymptotic will be used to refer to the slow adaptation regime  where $\mu\rightarrow 0$. Specifically, we will follow these steps:
\begin{itemize}
\item
We show that, in the stationary, steady-state regime, $\by_k(n)$ has a {\em limiting distribution} as $n$ goes to infinity (Theorem~1). 
\item
For small step-sizes, the steady-state distribution of $\by_k(n)$ approaches a Gaussian, i.e., it is {\em asymptotically normal} (Theorem~2). 
\item
We characterize the {\em large deviations} of the steady-state output $\by_k(n)$ in the slow adaptation regime when $\mu\rightarrow 0$ (Theorem~3). 
\item
The results of the above steps will provide a series of tools for designing the detector and characterizing its performance (Theorem~4).
\end{itemize}

\subsection{Comparison with Decaying Step-Size Solutions}
It is useful to contrast the above results with those pertaining to distributed detection algorithms with diminishing step-size~\cite{MouraLDGauss,MouraLDnonGauss,MouraLDnoisy}. The result in Theorem~1 reveals that, under stationary conditions, the detection statistic (i.e., the diffusion output $\by_k(n)$) converges to a limiting distribution, and the results in Theorem~2 add that such limiting distribution is approximately Gaussian in the slow adaptation regime. 
In contrast, in the diminishing step-size case, the detection statistic will collapse, as time elapses, into a {\em deterministic} value (e.g., the Kullback-Leibler divergence). Such convergence to a deterministic value reflects the continuously improving performance as time elapses, with diminishing step-sizes. In particular, under stationary conditions, the error probabilities for diminishing step-size algorithms decay exponentially {\em as functions of the time index $n$} --- see, e.g.~\cite{MouraLDGauss,MouraLDnonGauss,MouraLDnoisy}.
The latter feature must be contrasted with the results of our Theorems~3 and~4, where the exponential decay of the error probabilities does {\em not}  refer to the time index $n$. Instead, we find the new result that the error probabilities decay exponentially {\em as functions of the (inverse of the) step-size} $\mu$.

Finally, we would like to mention that the detailed statistical characterization offered by Theorems 1-3 is not confined to the specific detection problems we are dealing with. As a matter of fact, these results are of independent interest, and might be useful for the application of adaptive diffusion strategies in broader contexts.

\subsection{Main Results}
As explained in the previous section, we focus on a connected network of $S$ sensors, performing distributed detection by means of adaptive diffusion strategies.
The adaptive nature of the solution allows the network to track variations in the hypotheses being tested over time. 
In order to enable continuous adaptation and learning, we shall employ distributed diffusion strategies with a {\em constant} step-size parameter $\mu$. Now, let $\alpha_{k,\mu}$ and $\beta_{k,\mu}$ represent the steady-state  (as $n\rightarrow\infty$) Type-I and Type-II error probabilities at the $k$-th sensor. One of the main conclusions established in this paper can  
be summarized by the following scaling laws:
\beq
\boxed{
\alpha_{k,\mu}\stackrel{\cdot}{=}e^{-(1/\mu)\,S\,{\cal E}_0},\qquad
\beta_{k,\mu}\stackrel{\cdot}{=}e^{-(1/\mu)\,S\,{\cal E}_1}
}
\label{eq:mainres2}
\eeq
where the notation $\stackrel{\cdot}{=}$ means equality to the leading exponential order as $\mu$ goes to zero~\cite{CT}.
In the above expressions, the parameters  ${\cal E}_0$ and ${\cal E}_1$ are solely dependent on the moment generating function of the single-sensor data $\bx$, and of the decision regions. These parameters are {\em independent} of the step-size $\mu$, the number of sensors $S$, and the network connectivity. Result~(\ref{eq:mainres2}) has at least four important and insightful ramifications about the performance of adaptive schemes for distributed detection over networks.

To begin with, Eq.~(\ref{eq:mainres2}) reveals a  fundamental scaling law for distributed detection with diffusion adaptation, namely, it asserts that as the step-size decreases, the error probabilities are driven to zero exponentially as functions of $1/\mu$, and that the error exponents governing such a decay increase linearly in the number of sensors. These implications are even more revealing if examined in conjunction with the known results concerning the scaling law of the Mean-Square-Error (MSE) for adaptive distributed estimation over diffusion networks~\cite{ZhaoSayedLMSestimation,SayedSPmag}. 
Assuming a connected network with $S$ sensors, and using sufficiently small step-sizes $\mu\approx 0$, the MSE that is attained by sensor $k$ obeys (see expression (32) in~\cite{SayedSPmag}):
\beq
\textnormal{MSE}_k\; \propto \;\frac{\mu}{S},
\label{eq:MSEscales}
\eeq
where the symbol $\propto$ denotes proportionality. Some interesting symmetries are observed. In the estimation context, the MSE decreases as $\mu$ goes to zero, and the scaling rate improves linearly in the number of sensors.
Recalling that smaller values of $\mu$ mean a lower degree of adaptation, we observe that reaching a better inference quality costs in terms of adaptation speed. This is a well-known trade-off in the adaptive estimation literature between tracking speed and estimation accuracy.

\vspace*{5pt}
Second, we observe from~(\ref{eq:mainres2}) and~(\ref{eq:MSEscales}) that  the scaling laws governing errors of detection and estimation over distributed networks behave very differently, the former exhibiting an exponential decay proportional to $1/\mu$, while the latter is linear with decay proportional to $\mu$. 
The significance and elegance of this result for adaptive distributed networks lie in revealing an intriguing analogy with other more traditional inferential schemes.
As a first example, consider the standard case of a centralized, non-adaptive inferential system with $N$ i.i.d. data points. It is known that the error probabilities of the best detector decay exponentially fast to zero with $N$, while the optimal estimation error decays as $1/N$~\cite{shao,LehmannRomano}. 
Another important case is that of rate-constrained multi-terminal inference~\cite{TheCEOprob,BergerQuadCEO}. In this case the detection performance scales exponentially with the bit-rate $R$ while, again, the squared estimation error vanishes as $1/R$. 
Thus, at an abstract level, reducing the step-size corresponds to increasing the number of independent observations in the first system, or increasing the bit-rate in the second system.   
The above comparisons furnish an interesting interpretation for the step-size $\mu$ as the basic parameter quantifying the cost of information used by the network for inference purposes, much as the number of data $N$ or the bit-rate $R$ in the considered examples.

\vspace*{5pt}
A third aspect pertaining to the performance of the distributed network relates to the potential benefits of cooperation. These are already encoded into~(\ref{eq:mainres2}), and we have already implicitly  commented on them. Indeed, note that the error exponents increase linearly in the number of sensors. This implies that cooperation offers {\em exponential gains} in terms of detection performance. 

\vspace*{5pt}
The fourth and final ramification we would like to highlight relates to how much performance is lost by the {\em distributed} solution in comparison to a centralized stochastic gradient solution. 
Again, the answer is contained in~(\ref{eq:mainres2}). Specifically, the centralized solution is equivalent to a fully connected network, so that~(\ref{eq:mainres2}) applies to the centralized case as well. As already mentioned, the parameters ${\cal E}_0$ and ${\cal E}_1$ do not depend on the network connectivity, which therefore implies that, as the step-size $\mu$ decreases, the distributed diffusion solution of the inference problem exhibits a detection performance governed by the {\em same error exponents} of the centralized system. 
This is a remarkable conclusion and it is also consistent with results in the context of adaptive distributed estimation over diffusion networks~\cite{SayedSPmag}. 

We now move on to describe the adaptive distributed solution and to establish result~(\ref{eq:mainres2}) and the aforementioned properties.

\section{Existence of Steady-State Distribution}
\label{sec:steady}

Let $\by_n$ denote the $S\times 1$ vector that collects the state variables from across the network at time $n$, i.e., 
\beq
\by_n=\mbox{\rm col}\{\by_1(n),\,\by_2(n),\,\ldots,\by_S(n)\}.
\eeq
Likewise, we collect the local statistics $\{\bx_k(n)\}$ at time $n$ into the vector $\bx_n$.  It is then straightforward to verify from the diffusion strategy~(\ref{eq:diff1})--(\ref{eq:diff2}) that the vector $\by_n$ is given by:
\beq
\boxed
{
\by_n=(1-\mu)^n\,A^n \by_0  + \frac{\mu}{1-\mu} \sum_{i=1}^{n} (1-\mu)^{n-i+1} A^{n-i+1} \bx_i
}
\label{eq:vecATCfirst}
\eeq
We are concerned here with a {\em steady-state} analysis. Accordingly, we must examine the situation where the data are possibly nonstationary up to a certain time instant, after which they are drawn from the same stationary distribution for infinitely long time. 
This implies that, when performing the steady-state analysis, it suffices to assume that the data, for all $n\geq 1$, arise from one and the same distribution. The past history (including possible drifts occurred in the statistical conditions) that influences the overall algorithm evolution, is reflected in the initial state vector $\by_0$. 
In addition, since, for $n\geq 1$, we only need to specify the particular distribution from which data are drawn, in the forthcoming derivations we shall conduct our study with reference to a sequence of i.i.d. data with a given distribution. Later on, when applying the main findings to the detection problem, we shall use a subscript $h\in\{0,1\}$ to denote that data follow the distribution corresponding to a particular hypothesis.

We are now ready to show the existence and the specific shape of the limiting distribution.
By making the change of variables $i\leftarrow n-i+1$, Eq.~(\ref{eq:vecATCfirst}) can be written as 
\beq
\by_n=(1-\mu)^n\,A^n \by_0  + \frac{\mu}{1-\mu} \sum_{i=1}^{n} (1-\mu)^i A^i \bx_{n-i+1}.
\label{eq:changeofvariable}
\eeq
It follows that the state of the $k$-th sensor is given by:
\beqa
\by_k(n)&=&
\underbrace{(1-\mu)^n\,\sum_{\ell=1}^S b_{k,\ell} (n) \by_\ell(0)}_{\textnormal{transient}}
\nonumber\\
&+& 
\underbrace{
\frac{\mu}{1-\mu} 
\sum_{i=1}^{n} (1-\mu)^i  \sum_{\ell=1}^S b_{k,\ell}(i) \bx_\ell(n-i+1),
}_{\textnormal{steady-state}}\nonumber\\
\label{eq:mainATC}
\eeqa
where the scalars $b_{k,\ell}(n)$ are the entries of the matrix power:
\beq
B_n\dfz A^n.
\eeq
Since we are interested in reaching a {\em balanced} fusion of the observations, we shall assume that $A$ is {\em doubly stochastic} with second largest eigenvalue magnitude strictly less than one, which yields~\cite{SayedAcademicPress,xiao-boyd,Johnson-Horn}:
\beq
B_n\stackrel{n\rightarrow\infty}{\longrightarrow} \displaystyle{\frac 1 S}\,\mathds{1}\mathds{1}^T.
\label{eq:doublestoc}
\eeq

Now, we notice that the first term on the RHS of~(\ref{eq:mainATC}) vanishes almost surely (a.s.) (and, hence, in probability~\cite{shao}) with $n$, since, for any initial state vector $\by_0$, we have:
\beq
\left|(1-\mu)^n\,\sum_{\ell=1}^S b_{k,\ell} (n) \by_\ell(0)\right|\leq
(1-\mu)^n \,\sum_{\ell=1}^S |\by_\ell(0)|.
\eeq
Accordingly, if we are able to show that the second term on the RHS of~(\ref{eq:mainATC}) converges to a certain limiting distribution, we can then conclude that the variable $\by_k(n)$ converges as well to the same limiting distribution, as a direct application of Slutsky's Theorem~\cite{shao}.

In order to reveal the steady-state behavior of $\by_k(n)$, it suffices to focus on the last summation in~(\ref{eq:mainATC}).
We observe preliminarily that the term $\bx_{n-i+1}$ in~(\ref{eq:changeofvariable}) depends on the time index $n$ in such a way that the most recent datum $\bx_n$ is assigned the highest scaling weight, in compliance with the adaptive nature of the algorithm. However, since the vectors $\bx_i$ are i.i.d. across time, and since we shall be only concerned with the distribution of partial sums involving these terms, the statistical properties of the summation in~(\ref{eq:changeofvariable}) are left unchanged if we replace $\bx_{n-i+1}$ with a random vector $\bx^\prime_i$, where $\{\bx_i^\prime\}$ is a sequence of i.i.d. random vectors distributed similarly to the $\{\bx_{n-i+1}\}$. 
Formally, as regards the steady-state term on the RHS of~(\ref{eq:mainATC}), we can write:
\beqa
\lefteqn{
\frac{\mu}{1-\mu} 
\sum_{i=1}^{n} (1-\mu)^i  \sum_{\ell=1}^S b_{k,\ell}(i) \bx_\ell(n-i+1)
}
\nonumber\\
&\stackrel{d}{=}&
\frac{\mu}{1-\mu} 
\sum_{i=1}^{n} (1-\mu)^i  \sum_{\ell=1}^S b_{k,\ell}(i) \bx^{\prime}_\ell(i)
\dfz
\sum_{i=1}^{n} \bz_k(i),
\nonumber\\
\label{eq:sumexpress}
\eeqa
where $\stackrel{d}{=}$ denotes equality {\em in distribution}, and where the definition of $\bz_k(i)$ should be clear. 
As a result, we are faced with a sum of independent, but {\em not} identically distributed, random variables.
Let us evaluate the first two moments of the sum:
\beq
\E \left(\sum_{i=1}^n \bz_k(i)\right)
=
\E \bx \, \sum_{i=1}^n \mu(1-\mu)^{i-1}  \underbrace{\sum_{\ell=1}^S b_{k,\ell}(i)}_{=1}
\stackrel{n\rightarrow\infty}{\longrightarrow}
\E \bx,
\label{eq:oneseries}
\eeq
and 
\beqa
\VAR \left( \sum_{i=1}^n \bz_k(i) \right)&=& 
\sigma^2_x \, \sum_{i=1}^n \mu^2(1-\mu)^{2(i-1)} \underbrace{\sum_{\ell=1}^S  b^2_{k,\ell}(i)}_{ \leq 1}
\nonumber\\
&\leq& \frac{\sigma^2_x  \,\mu}{2-\mu}<\infty,
\label{eq:twoseries}
\eeqa
where $\VAR$ denotes the variance operator, and $\sigma^2_x\dfz\VAR(\bx)$.
We have thus shown that the expectation of the sum expression from~(\ref{eq:sumexpress}) converges to $\E \bx$, and that its variance converges to a finite value.
In view of the Infinite Convolution Theorem --- see~\cite[p. 266]{FellerBookV2}, these two conditions are sufficient to conclude that the  RHS of~(\ref{eq:sumexpress}), i.e., the sum of random variables $\bz_k(i)$, converges in distribution as $n\rightarrow\infty$, and the first two moments of the limiting distribution are equal to $\E\bx$ and $\sum_{i=1}^\infty \VAR(\bz_k(i))$.
The random variable characterized by the limiting distribution will be denoted by $\by_{k,\mu}^{\star}$, where we make explicit the dependence upon the step-size $\mu$ for later use.

The above statement can be sharpened to ascertain that the sum of random variables $\bz_k(i)$ actually converges almost surely. This conclusion can be obtained by applying Kolmogorov's Two Series Theorem~\cite{FellerBookV2}. 
In view of the a.s. convergence, it makes sense to define the limiting random variable $\by_{k,\mu}^\star$ as:
\beq
\boxed{
\by^\star_{k,\mu}\dfz 
\sum_{i=1}^{\infty} \sum_{\ell=1}^S 
\mu\,(1-\mu)^{i-1} b_{k,\ell}(i) \bx^{\prime}_\ell(i)
}
\label{eq:zstardef}
\eeq
We wish to avoid confusion here. We are not stating that the actual diffusion output $\by_k(n)$ converges almost surely (a behavior that would go against the adaptive nature of the diffusion algorithm). 
We are instead claiming that $\by_k(n)$ converges in distribution to a random variable $\by_{k,\mu}^\star$ that can be conveniently defined in terms of the a.s. limit~(\ref{eq:zstardef}).

The main result about the steady-state behavior of the diffusion output is summarized below (the symbol $\rightsquigarrow$ means convergence in distribution).

\vspace*{5pt}
\noindent
\textsc{Theorem 1:} 
{\em 
(Steady-state distribution of $\by_k(n)$). 
The state variable $\by_k(n)$ that is generated by the diffusion strategy~(\ref{eq:diff1})--(\ref{eq:diff2}) is asymptotically stable in distribution, namely,
\beq
\boxed{
\by_k(n)\stackrel{n\rightarrow\infty}{\rightsquigarrow} \by^\star_{k,\mu}
}
\label{eq:Theo1}
\eeq
}
~\hfill$\square$

\vspace*{5pt}
\noindent
It is useful to make explicit the meaning of Theorem~1. By definition of convergence in distribution (or weak convergence), the result~(\ref{eq:Theo1}) can be formally stated as~\cite{billingsley-book,LehmannRomano}:
\beq
\lim_{n\rightarrow\infty} \P[\by_k(n)\in\Gamma]=\P[\by^\star_{k,\mu} \in \Gamma],
\label{eq:weakconv}
\eeq
for any set $\Gamma$ such that $\P[\by^\star_{k,\mu} \in \partial\Gamma]=0$, where $\partial\Gamma$ denotes the boundary of $\Gamma$.
It is thus seen that the properties of the steady-state variable $\by^\star_{k,\mu}$ will play a key role in determining the steady-state performance of the diffusion output. 
Accordingly, we state two useful properties of $\by^\star_{k,\mu}$.

First, when the local statistic $\bx_k(n)$ has an {\em absolutely continuous} distribution (where the reference measure is the Lebesgue measure over the real line), it is easily verified that the distribution of $\by^\star_{k,\mu}$ is {\em absolutely continuous as well}. Indeed, note that we can write $\by^\star_{k,\mu}=\bz_k(1)+\sum_{i=2}^\infty \bz_k(i)$. Now observe that $\bz_k(1)$, which has an absolutely continuous distribution by assumption, is independent of the other term. The result follows by the properties of convolution and from the fact that the distribution of the sum of two independent variables is the convolution of their respective distributions.

Second, when the local statistic $\bx_k(n)$ is a {\em discrete} random variable, by the Jessen-Wintner law~\cite{JessenWintner,breiman1968probability}, we can only conclude that $\by^\star_{k,\mu}$ is of {\em pure type}, namely, its distribution is pure: absolutely continuous, or discrete, or continuous but singular.

An intriguing case is that of the so-called {\em Bernoulli convolutions}, i.e.,  random variables of the form $\sum_{i=1}^\infty (1-\mu)^{i-1} \bx(i)$, where $\bx(i)$ are equiprobable $\pm 1$. For this case, it is known that if $1/2<\mu<1$, then the limiting distribution is a {\em Cantor} distribution~\cite{ErdosBernoulliConv}. 
This is an example of a distribution that is neither discrete nor absolutely continuous. When $\mu<1/2$, which is relevant for our discussion since we shall be concerned with small step-sizes, the situation is markedly different, and the  distribution is absolutely continuous for almost all values of $\mu$. 

Before proceeding, we stress that we have proved that a steady-state distribution for $\by_k(n)$ exists, but its form is not known. Accordingly, even in steady-state, the structure of the optimal test is still unknown. 
In tackling this issue, and recalling that the regime of interest is that of slow adaptation, we now focus on the case $\mu\ll 1$.

\section{The Small-$\mu$ Regime.}
\label{sec:smallmu}

While the exact form of the steady-state distribution is generally impossible to evaluate, it is nevertheless possible to approximate it well for small values of the step-size parameter. 
Indeed, in this section we prove two results concerning the statistical characterization of the steady-state distribution for $\mu\rightarrow 0$.
The first one is a result of {\em asymptotic normality}, stating that  $\by^\star_{k,\mu}$ approaches a Gaussian random variable with known moments as $\mu$ goes to zero (Theorem~2). 
The second finding (Theorem~3) provides the complete characterization for the {\em large deviations} of $\by^\star_{k,\mu}$. 
In the following, ${\cal N}(a,b)$ is a shortcut for a Gaussian distribution with mean $a$ and variance $b$, and the symbol $\sim$ means ``distributed as".

\vspace*{5pt}
\noindent
\textsc{Theorem 2:}
{\em (Asymptotic normality of $\by^\star_{k,\mu}$ as $\mu\rightarrow 0$). 
Under the assumption $\E|\bx_k(n)|^3<\infty$, the variable $\by^\star_{k,\mu}$ fulfills, for all $k=1,2,\dots,S$:}
\beq
\boxed
{
\frac{\by^\star_{k,\mu}-\E \bx}{\sqrt{\mu}}\stackrel{\mu\rightarrow 0}{\rightsquigarrow} {\cal N}\left(0,\frac{\sigma^2_x}{2\,S}\right)
}
\label{eq:CLT}
\eeq

\vspace*{5pt}
\noindent
{\em Proof:} The argument requires dealing with independent but non-identically distributed random variables, as done in the Lindeberg-Feller CLT (Central Limit Theorem)~\cite{FellerBookV2}. This theorem, however, does not apply to our setting since the asymptotic parameter is {\em not} the number of samples, but rather the step-size. Some additional effort is needed, and the detailed technical derivation is deferred to Appendix~A.
~\hfill$\square$ 

\subsection{Implications of Asymptotic Normality}
Let us now briefly comment on several useful implications that follow from the above theorem: 
\begin{enumerate}
\item
First, note that {\em all sensors} share, for $\mu$ small enough, the {\em same} distribution, namely, the inferential diffusion strategy equalizes the statistical behavior of the agents.  This finding complements well results from~\cite{ZhaoSayedLMSestimation,SayedSPmag,ChenSayedGlobalSIP} where the asymptotic equivalence among the sensors has been proven in the context of mean-square-error estimation. One of the main differences between the estimation context and the detection context studied in this article is that in the latter case, the regression data is deterministic  and the randomness arises from the stochastic nature of the statistics $\{\bx_k(n)\}$. For this reason, the steady-state distribution in~(\ref{eq:CLT}) is characterized in terms of the moments of these statistics and not in terms of the moments of regression data, as is the case in the estimation context. 
\item
The result of Theorem~2 is valid provided that the connectivity matrix fulfills~(\ref{eq:doublestoc}). This condition is satisfied when the network topology is strongly-connected, i.e., there exists a path with nonzero weights connecting any two arbitrary nodes and at least one node has $a_{k,k}>0$~\cite{SayedAcademicPress}. Obviously, condition~(\ref{eq:doublestoc}) is also satisfied in the fully connected case when $a_{k,\ell}=b_{k,\ell}=1/S$ for all $k,\ell=1,2,\dots,S$. This latter situation would correspond to a representation of  the centralized stochastic gradient algorithm, namely, an implementation of the form
\beq
\by^{(c)}(n)=\by^{(c)}(n-1)+\frac{\mu}{S}\sum_{\ell=1}^S [\bx_{\ell}(n)-\by^{(c)}(n-1)],  
\label{eq:centstochalg}
\eeq
where $\by^{(c)}(n)$ denotes the output by the centralized solution at time $n$. The above algorithm can be deduced from~(\ref{eq:diff1})--(\ref{eq:diff2}) by defining 
\beq
\by^{(c)}(n)\dfz \frac{1}{S}\sum_{\ell=1}^S \by_{\ell}(n).
\eeq
Now, since the moments of the limiting Gaussian distribution in~(\ref{eq:CLT}) are independent of the particular connectivity matrix, the net effect is that each agent of the {\em distributed} network acts, asymptotically, as the {\em centralized} system. 
This result again complements well results in the estimation context where the role of the statistics variables $\{\bx_{k}(n)\}$ is replaced by that of stochastic regression data~\cite{ZhaoSayedICASSP2013}.
\item
The asymptotic normality result is powerful in approximating the steady-state distribution for relatively small step-sizes, thus  enabling the analysis and design of inferential diffusion networks in many different contexts. 
With specific reference to the detection application that is the main focus here, Eq.~(\ref{eq:CLT}) can be exploited for an accurate threshold setting when one desires to keep under control one of the two errors, say, the false-alarm probability, as happens, e.g., in the Neyman-Pearson setting~\cite{LehmannRomano}.   
To show a concrete example on how this can be done, let us assume that, without loss of generality, $\E_0\bx<\E_1\bx$, and consider a single-threshold detector for which: 
\beq
\Gamma_0=\{\gamma \in\mathbb{R}:\,
\gamma\leq \eta_\mu
\},\qquad \Gamma_1=\mathbb{R}\setminus\Gamma_0,
\label{eq:gamma0}
\eeq
where the threshold is set as
\beq
\eta_\mu=\E_0\bx + \sqrt{\frac{\mu \sigma_{x,0}^2}{2 S}}\,Q^{-1}(\bar\alpha).
\label{eq:NPth1}
\eeq
Here, $\sigma^2_{x,0}$ is the variance of $\bx$ under ${\cal H}_0$, $Q(\cdot)$ denotes the complementary CDF for a standard normal distribution, and $\bar\alpha$ is the prescribed false-alarm level.
By~(\ref{eq:CLT}), it is straightforward to check that this threshold choice ensures
\beq
\lim_{\mu\rightarrow 0} \P_0[\by_{k,\mu}^\star>\eta_\mu]=\bar\alpha.
\label{eq:NPpfa1}
\eeq
\end{enumerate}

In summary, Theorem~2 provides an approximation of the diffusion output distribution for small step-sizes.
At first glance, this may seem enough to obtain a complete characterization of the detection problem. A closer inspection reveals that this is not the case. A good example to understand why Theorem~2 alone is insufficient for characterizing the detection performance is obtained by examining the Neyman-Pearson threshold setting just described in~(\ref{eq:NPth1})--(\ref{eq:NPpfa1}) above. 
While we have seen that the asymptotic behavior of the false-alarm probability in~(\ref{eq:NPpfa1}) is completely determined by the application of Theorem~2, the situation is markedly different as regards the miss-detection probability $\P_1[\by_{k,\mu}^\star\leq\eta_\mu]$. 
Indeed, by using~(\ref{eq:NPth1}) we can write:
\beqa
\lefteqn{
\P_1[\by_{k,\mu}^\star\leq\eta_\mu]
=
\P_1\left[
\frac{\by_{k,\mu}^\star-\E_1\bx}{\sqrt{\mu}}\leq\frac{\eta_\mu-\E_1\bx}{\sqrt{\mu}}
\right]
}
\nonumber\\
&=&\P_1\left[
\frac{\by_{k,\mu}^\star-\E_1\bx}{\sqrt{\mu}}\leq \frac{\E_0\bx-\E_1\bx}{\sqrt{\mu}}+\sqrt{\frac{\sigma_{x,0}^2}{2 S}}\,Q^{-1}(\bar\alpha)
\right].\nonumber\\
\eeqa
Since $\E_0\bx < \E_1 \bx$, the quantity $\frac{\E_0\bx-\E_1\bx}{\sqrt{\mu}}$ diverges to $-\infty$ as $\mu\rightarrow 0$. As a consequence, the fact that $\frac{\by_{k,\mu}^\star-\E_1\bx}{\sqrt{\mu}}$ is asymptotically normal does not provide much more insight than revealing that the miss-detection probability converges to zero as $\mu\rightarrow 0$. A meaningful asymptotic analysis would instead require to examine the way this convergence takes place (i.e., the error exponent).   
The same kind of problem is found when one lets {\em both} error probabilities vanish exponentially, such that the Type-I and Type-II detection error exponents furnish a meaningful asymptotic characterization of the detector.
In order to fill these gaps, the study of the {\em large} deviations of $\by_{k,\mu}^\star$ is needed.

\subsection{Large Deviations of $\by^\star_{k,\mu}$.}
\label{subsec:LDP}
From~(\ref{eq:CLT}) we learn that, as $\mu\rightarrow 0$, the diffusion output shrinks down to its limiting expectation $\E\bx$ and that the {\em small} (of order $\sqrt{\mu}$)  deviations around this value have a Gaussian shape. 
But this conclusion is not helpful when working with {\em large} deviations, namely, with terms like:
\beq
\P[|\by^\star_{k,\mu}-\E\bx|>\delta]
\stackrel{\mu\rightarrow 0}{\longrightarrow} 0,\quad \delta>0,
\label{eq:Large}
\eeq
which play a significant role in detection applications.
While the above convergence to zero can be inferred from~(\ref{eq:CLT}), it is well known that~(\ref{eq:CLT}) is not sufficient in general to obtain the rate at which the above probability vanishes. 
In order to perform accurate design and characterization of reliable inference systems~\cite{Dembo-Zeitouni,DenHollander} it is critical to assess this rate of convergence, which turns out to be the main purpose of a large deviations analysis. 

Accordingly, we will be showing in the sequel that the process $\by^\star_{k,\mu}$ obeys a Large Deviations Principle (LDP), namely, that the following limit exists~\cite{Dembo-Zeitouni,DenHollander}:
\beq
\lim_{\mu\rightarrow 0} \mu\, \ln  \P[\by^\star_{k,\mu}\in \Gamma]=
-\inf_{\gamma\in \Gamma} I(\gamma)\dfz - I_\Gamma,
\label{eq:LDPdef}
\eeq
for some $I(\gamma)$ that is called the {\em rate function}. 
Equivalently:
\beq
\P[\by^\star_{k,\mu}\in\Gamma]=e^{-(1/\mu)\,I_\Gamma + o(1/\mu)}
\stackrel{\cdot}{=}e^{-(1/\mu)\,I_\Gamma},
\label{eq:expoexpo}
\eeq
where $o(1/\mu)$ stands for any correction term growing slower than $1/\mu$, namely, such that $\mu\,o(1/\mu)\rightarrow 0$ as $\mu\rightarrow 0$, and the notation $\stackrel{\cdot}{=}$ was introduced in~(\ref{eq:mainres2}).
From~(\ref{eq:expoexpo}) we see that, in the large deviations framework, only the dominant exponential term is retained, while discarding any sub-exponential terms. It is also interesting to note that, according to~(\ref{eq:expoexpo}), the probability that $\by^\star_{k,\mu}$ belongs to a given region $\Gamma$ is dominated by the infimum $I_\Gamma$ of the rate function $I(\gamma)$ within the region $\Gamma$. In other words, the smallest exponent ($\Rightarrow$ highest probability) dominates, which is well explained in~\cite{DenHollander} through the statement: ``{\em any large deviation is done in the least unlikely of all the unlikely ways}''. 

In summary, the LDP generally implies an exponential scaling law for probabilities, with an exponent governed by the rate function.
Therefore, knowledge of the rate function is enough to characterize the exponent in~(\ref{eq:expoexpo}). We shall determine the expression for $I(\gamma)$ pertinent to our problem in Theorem~3 further ahead --- see Eq.~(\ref{eq:ratefunSomega}).

In the traditional case where the statistic under consideration is the arithmetic average of i.i.d. data, the asymptotic parameter is the number of samples and the usual tool for determining the rate function in the LDP is Cram\'er's Theorem~\cite{Dembo-Zeitouni,DenHollander}. Unfortunately, in our adaptive and distributed setting, we are dealing with a more general  statistic $\by^\star_{k,\mu}$, whose dependence is on the step-size parameter and not on the number of samples. 
Cram\'er's Theorem is not applicable in this case, and we must resort to a more powerful tool, known as the G\"artner-Ellis Theorem~\cite{Dembo-Zeitouni,DenHollander}, stated below in a form that uses directly the set of assumptions relevant for our purposes.

\vspace*{5pt}
\noindent
\textsc{G\"artner-Ellis Theorem~\cite{DenHollander}}. 
{\em Let $\bz_\mu$ be a family of random variables with Logarithmic Moment Generating Function (LMGF)
$\phi_{\mu}(t)=\ln \E \exp\{t \bz_\mu\}.$
If 
\beq
\phi(t)\dfz\lim_{\mu\rightarrow 0} \mu\, \phi_{\mu}(t/\mu)
\eeq
exists, with $\phi(t)<\infty$ for all $t\in\mathbb{R}$, and $\phi(t)$ is differentiable in $\mathbb{R}$, then $\bz_\mu$ satisfies the LDP property~(\ref{eq:LDPdef}) with rate function given by the Fenchel-Legendre transform of $\phi(t)$, namely:
\beq
\Phi(\gamma)\dfz\sup_{t\in\mathbb{R}}[\gamma t -\phi(t)].
\label{eq:FLtransf}
\eeq
}
~\hfill$\square$

\vspace*{5pt}
\noindent
In what follows, we shall use capital letters to denote Fenchel-Legendre transforms, as done in~(\ref{eq:FLtransf}).

We now show how the  result allows us to assess the asymptotic performance of the diffusion output in the inferential network.
Let us introduce the LMGF of the data $\bx_k(n)$, and that of the steady-state variable $\by^\star_{k,\mu}$, respectively:
\beqa
\psi(t)&\dfz& \ln \E \exp\{t \bx_k(n)\},\\
\phi_{k,\mu}(t)&\dfz& \ln \E \exp\{t \by_{k,\mu}^\star\}.
\eeqa

\vspace*{5pt}
\noindent
\textsc{Theorem 3:} 
{\em 
(Large deviations of  $\by^\star_{k,\mu}$ as $\mu\rightarrow 0$). 
Assume that $\psi(t)<\infty$ for all $t\in \mathbb{R}$. 
Then, for all $k=1,2,\dots,S$:
\begin{itemize}
\item[$i)$]
\beq
\boxed
{
\phi(t)\dfz\lim_{\mu\rightarrow 0} \mu\,
\phi_{k,\mu}(t/\mu)
=
S\,
\omega(t/S)
}
\label{eq:LMGFlimit}
\eeq
where
\beq
\boxed
{
\omega(t)\dfz\int_{0}^{t}\frac{\psi(\tau)}{\tau}d\tau
\label{eq:omegadef}
}
\eeq
\item[$ii)$]
The steady-state variable $\by^\star_{k,\mu}$ obeys the LDP with a rate function given by:
\beq
\boxed
{
I(\gamma)=S\,\Omega(\gamma)
}
\label{eq:ratefunSomega}
\eeq
that is, by the Fenchel-Legendre transform of $\omega(t)$ multiplied by the number of sensors $S$.
\end{itemize}
}

\vspace*{5pt}
\noindent
{\em Proof:} See Appendix B. 
~\hfill$\square$

\subsection{Main Implications of Theorem 3}
\label{subsec:mainimplic}

From Theorem 3, a number of interesting conclusions can be drawn:
\begin{itemize}
\item
The function $\omega(t)$ in~(\ref{eq:omegadef}) depends only upon the LMGF $\psi(t)$ of the original statistic $\bx_k(n)$, and does {\em not} depend on the number of sensors.
\item
As a consequence of the above observation, part $ii)$ implies that the rate function (and, therefore, the large deviations exponent) of the diffusion output depends {\em linearly on the number of sensors}. 
Moreover, the rate can be determined by knowing only the statistical distribution of the input data $\bx_k(n)$.
\item
The rate function does not depend on the particular sensor $k$. This implies that {\em all sensors are asymptotically equivalent also in terms of large deviations}, thus strengthening what we have already found in terms of asymptotic normality --- see Theorem 2 and the subsequent discussion.
\item
Theorem 3 can be applied to the centralized stochastic algorithm~(\ref{eq:centstochalg}) as well, and, again, the diffusion strategy is able to match, asymptotically, the {\em centralized} solution.
\end{itemize}

Before ending this section, it is useful to comment on some essential features of the rate function $\Omega(\gamma)$, which will provide insights on its usage in connection with the distributed detection problem. 
To this aim, we refer to the following convexity properties shown in Appendix C (see also~\cite{Dembo-Zeitouni}, Ex. 2.2.24, and~\cite{DenHollander}, Ex. I.16):
\begin{itemize}
\item[$i)$]
$\omega^{\prime\prime}(t)>0$ for all $t\in\mathbb{R}$, implying that $\omega(t)$ is strictly convex.
\item[$ii)$]
$\Omega(\gamma)$ is strictly convex in the interior of the set:
\beq
{\cal D}_{\Omega}=\{\gamma\in\mathbb{R}:\;\Omega(\gamma)<\infty\}.
\eeq
\item[$iii)$]
$\Omega(\gamma)$ attains its unique minimum at $\gamma=\E\bx$, with
\beq
\Omega(\E\bx)=0.
\eeq
\end{itemize}

\begin{figure*}
\centerline{\includegraphics[width=.8\textheight]{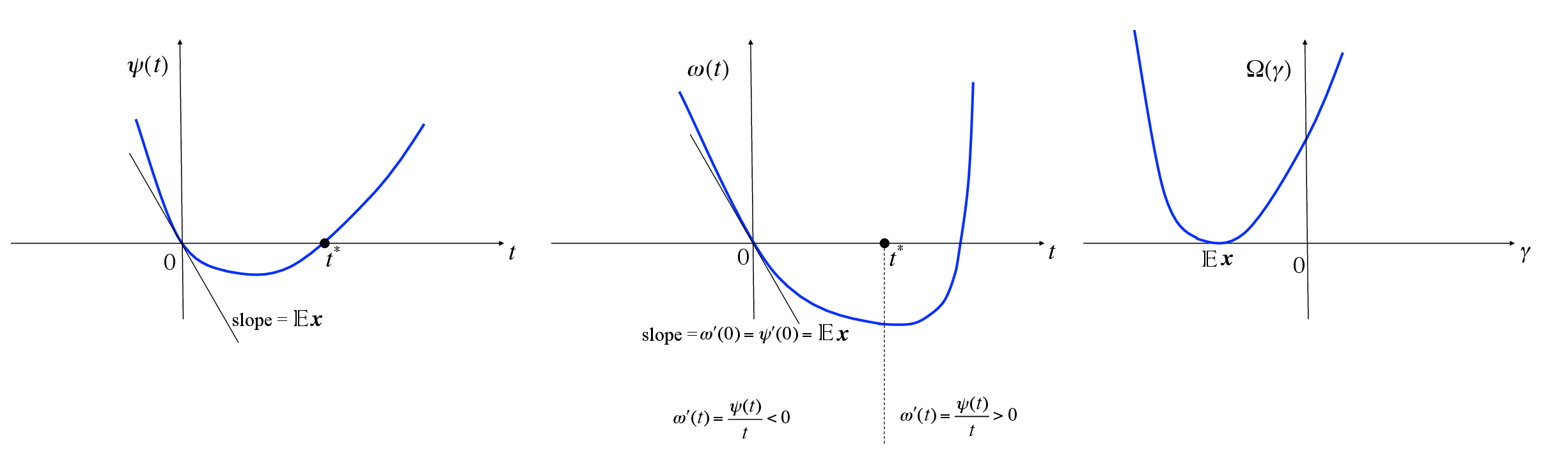}}
\caption{Leftmost panel: The LMGF $\psi(t)$ of the original data $\bx_k(n)$; its slope at the origin is $\E\bx$. Middle panel: The function $\omega(t)$ defined by~(\ref{eq:omegadef}) is strictly convex; its slope at the orgin is also equal to $\E\bx$. The labels underneath the plot illustrate the intervals over which $\omega^\prime(t)$ is negative and positive  for the LMGF $\psi(t)$ shown in the leftmost plot. Rightmost panel: The Fenchel-Legendre transform, $\Omega(\gamma)$, which is relevant for the evaluation of the rate function, attains the minimum value of zero at $\gamma=\E\bx$.}
\label{fig:RateFunGen}
\end{figure*}

\noindent
In light of these properties, it is possible to provide a geometric interpretation for the main quantities in Theorem~3, as illustrated in Fig.~\ref{fig:RateFunGen}.
The leftmost panel shows a typical behavior of the LMGF of the original data $\bx_k(n)$. 
Using the result $\omega^\prime(t)=\psi(t)/t$, and examining the sign of $\psi(t)/t$, it is possible to deduce the corresponding typical behavior of $\omega(t)$, depicted in the middle panel. As it can be seen, the slope at the origin is preserved, and is still equal to the expectation of the original data, $\E\bx$. The intersection with the $t$-axis is changed, and moves further to the right in the considered example.
Starting from $\omega(t)$, it is possible to draw a sketch of its Fenchel-Legendre transform $\Omega(\gamma)$ (rightmost panel), which illustrates its convexity properties, and the fact that the minimum value of zero is attained only at $\gamma=\E\bx$.

\section{The Distributed Detection Problem}
\label{sec:detect}

The tools and results developed so far allow us to address in some detail the detection problem we are interested in.
Let us denote the decision regions in favor of ${\cal H}_0$ and ${\cal H}_1$ by $\Gamma_0$ and $\Gamma_1$, respectively. We assume that they are the same at all sensors because, in view of the asymptotic equivalence among  sensors proved in the previous section, there is no particular interest in making a different choice. Note, however, that all the subsequent development does not rely on this assumption and applies, {\em mutatis mutandis}, to the case of distinct decision regions used by distinct agents.
 
The Type-I and Type-II error probabilities at the $k$-th sensor at time $n$ are defined in~(\ref{eq:alphaknfirstdef}) and~(\ref{eq:betaknfirstdef}), respectively.
Since we are interested in their {\em steady-state} behavior, namely, for an increasingly large interval where a certain hypothesis stays in force, the only distribution that matters is that corresponding to such hypothesis. Therefore, it is legitimate to write:
\beqa
\lim_{n\rightarrow\infty} \alpha_k(n)&=&\lim_{n\rightarrow\infty}\P_0[\by_k(n)\in \Gamma_1 ],\\
\lim_{n\rightarrow\infty} \beta_k(n)&=&\lim_{n\rightarrow\infty}\P_1[\by_k(n)\in \Gamma_0 ],
\eeqa
where the subscripts $0$ and $1$ denote here the (stationary) situation where the data collected for all $n\geq 1$ come from one and the same distribution. As already observed, this simply corresponds to saying that the stationarity period used to compute the steady-state distribution starts at time $n=1$. 
Some questions arise. Do these limits exist? Do these probabilities vanish as $n$ approaches infinity?
Theorem 1 provides the answers.  Indeed, we found that $\by_k(n)$ stabilizes in distribution as $n$ goes to infinity. In the sequel, in order to avoid dealing with pathological cases, we shall assume that $\P_0[\by^\star_{k,\mu}\in\partial\Gamma_1]=0$ and that $\P_1[\by^\star_{k,\mu}\in\partial\Gamma_0]=0$. This is a mild assumption, which is verified, for instance, when the limiting random variable $\by_{k,\mu}^\star$ has an absolutely continuous distribution, and the decision regions are not so convoluted to have boundaries with strictly positive measure. 
Accordingly, by invoking the weak convergence result of Theorem~1, and in view of~(\ref{eq:weakconv}) we can write: 
\beqa
\alpha_{k,\mu}&\dfz& \lim_{n\rightarrow\infty} \alpha_k(n)=\P_0[\by^\star_{k,\mu} \in \Gamma_1],\label{eq:alphabet1}\\
\beta_{k,\mu}&\dfz& \lim_{n\rightarrow\infty} \beta_k(n)=\P_1[\by^\star_{k,\mu} \in \Gamma_0],
\label{eq:alphabet2}
\eeqa
where the dependence upon $\mu$ has been made explicit for later use.  
We notice that, in the above, we work with decision regions that do not depend on $n$, which corresponds exactly to the setup of Theorem 1. Generalizations where the regions are allowed to change with $n$ can be handled by resorting to known results from asymptotic statistics.
To give an example, consider the meaningful case of a detector with a sequence of thresholds $\eta(n)$ that converges to a value $\eta$ as $n\rightarrow\infty$. Here,
\beq
\lim_{n\rightarrow\infty}\P_h[\by_k(n) > \eta(n)]=\P_h[\by^\star_{k,\mu} > \eta],
\eeq
which can be seen, e.g., as an application of Slutsky's Theorem~\cite{shao,LehmannRomano}.

From~(\ref{eq:alphabet1})--(\ref{eq:alphabet2}), it turns out that, as time elapses, the error probablities do not vanish exponentially. As a matter of fact, they do not vanish at all. This situation is in contrast to what happens in the case of running consensus strategies  with diminishing step-size  studied in the literature~\cite{running-cons,asymptotic-rc,Bracaetal-Pageconsensus,MouraLDGauss,MouraLDnonGauss,MouraLDnoisy}. 
We wish to avoid confusion here. In the diminishing step-size case, one does need to examine the effect of large deviations~\cite{MouraLDGauss,MouraLDnonGauss,MouraLDnoisy}  for large $n$, quantifying the rate of decay to zero of the error probabilities {\em as time progresses}. 
In the adaptive context, on the other hand, where {\em constant} step-sizes are used to enable continuous adaptation and learning, the large deviations analysis is totally different, in that it is aimed at characterizing the decaying rate of the error probabilities {\em as the step-size $\mu$ approaches zero.} 

Returning to the detection performance evaluation~(\ref{eq:alphabet1})--(\ref{eq:alphabet2}), we stress that the steady-state values of these error probabilities are unknown, since the distribution of $\by^\star_{k,\mu}$ is generally unknown. However, the large deviations result offered by Theorem 3 allows us to characterize the {\em error exponents} in the regime of small step-sizes. 

Theorem 3 can be tailored to our detection setup as follows (subscripts $0$ and $1$ are used to indicate that the statistical quantities are evaluated under ${\cal H}_0$ and ${\cal H}_1$, respectively):

\vspace*{5pt}
\noindent
{\em 
\textsc{Theorem 4:} (Detection error exponents). For $h\in\{0,1\}$, let $\Gamma_h$ be the decision regions --independent of $\mu$-- and assume that $\psi_h(t)<\infty$ for all $t\in\mathbb{R}$, and define:
\beq
\omega_h(t)\dfz\int_{0}^{t}\frac{\psi_h(\tau)}{\tau}d\tau.
\label{eq:Omegah}
\eeq
Then, for all $k=1,2,\dots,S$, Eq.(\ref{eq:mainres2}) holds true, namely,
\beq
\boxed{
\lim_{\mu\rightarrow 0} \mu\, \ln\alpha_{k,\mu}=-S\,{\cal E}_0,
\qquad
\lim_{\mu\rightarrow 0} \mu\, \ln\beta_{k,\mu}=-S\,{\cal E}_1
}
\label{eq:detperf}
\eeq
with
\beq
\boxed
{
{\cal E}_0=\inf_{\gamma\in \Gamma_1} \Omega_0(\gamma),\qquad
{\cal E}_1=\inf_{\gamma\in \Gamma_0} \Omega_1(\gamma)
}
\label{eq:E01}
\eeq
where $\Omega_h(\gamma)$ is the Fenchel-Legendre transform of $\omega_h(t)$.
}
~\hfill$\square$ 

\vspace*{10pt}
\noindent
\textsc{Remark I}. The technical requirement that the LMGFs $\psi_0(t)$ and $\psi_1(t)$ are finite is met in many practical detection problems, as already shown in~\cite{MouraLDnonGauss}. In particular, the assumption is clearly verified when the observations have (the same, under the two hypotheses) compact support, a special interesting case being that of discrete variables supported on a finite alphabet; and for shift-in-mean detection problems where the data distributions fulfill mild regularity conditions --- see Remark II in~\cite{MouraLDnonGauss} for a detailed list.

\begin{figure}[t]
\centerline{\includegraphics[width=.45\textheight]{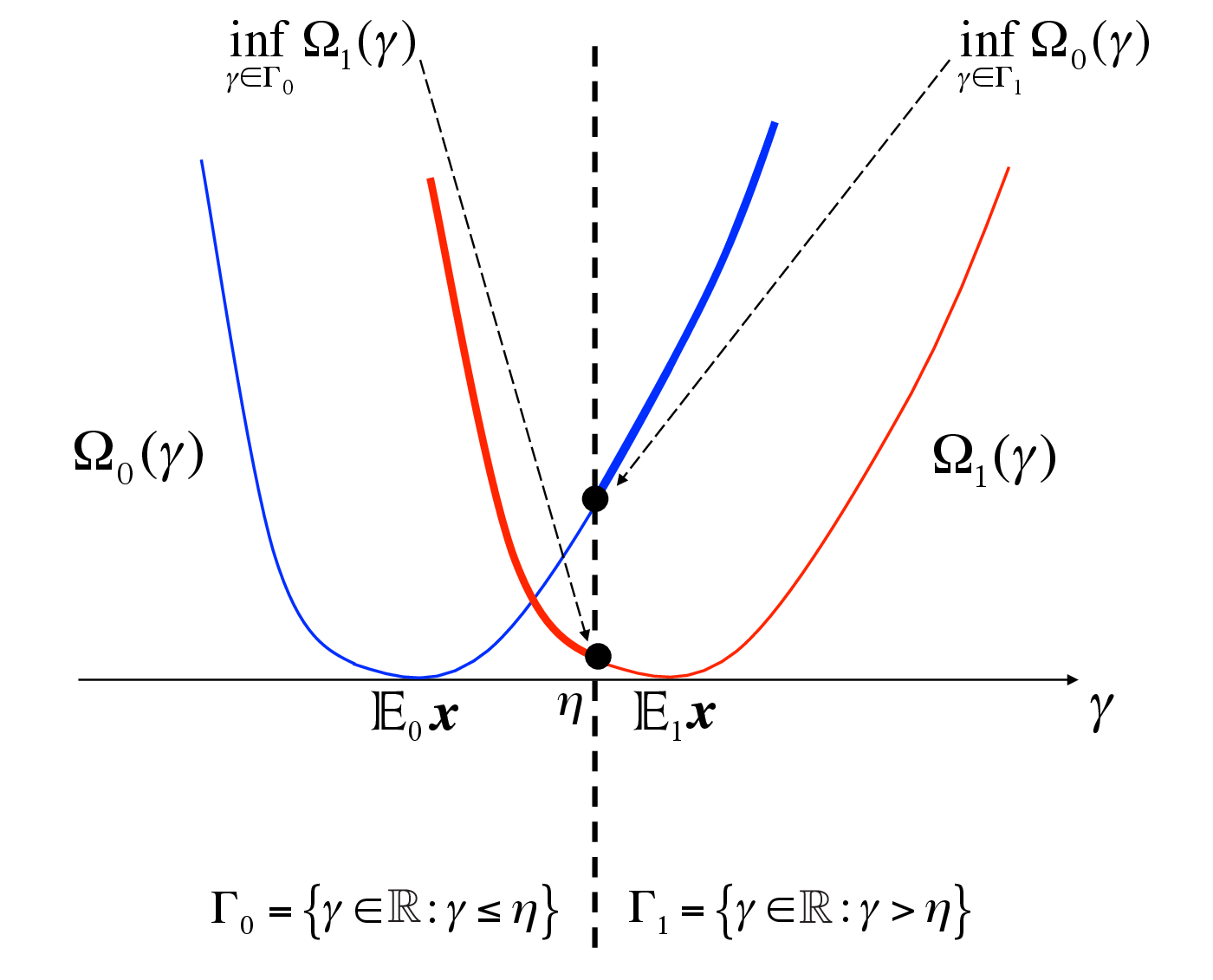}}
\caption{A geometric view of Theorem~4.}
\label{fig:Theo4Geo}
\end{figure}

\vspace*{5pt}
\noindent
\textsc{Remark II}. As typical in large deviations analysis, we have worked with regions $\Gamma_0$ and $\Gamma_1$ that do not depend on the step-size $\mu$. Generalizations are possible to the case in which these regions depend on $\mu$.  A relevant case where this might be useful is the Neyman-Pearson setup, where one needs to work with a fixed (non-vanishing) value of the false-alarm probability. 
An example of this scenario is provided in Sec.~\ref{subsec:Laplace} --- see the discussion following~(\ref{eq:betamuNP}) --- along with the detailed procedure for the required generalization. 

\vspace*{5pt}
In Fig.~\ref{fig:Theo4Geo}, we provide a geometric interpretation that can be useful to visualize the main message conveyed by Theorem~4.
In order to rule out trivial cases, we assume that $\E_0\bx\neq\E_1\bx$, as happens, e.g., in the standard situation where the local statistic $\bx_k(n)$ is a log-likelihood ratio and the detection problem is identifiable~\cite{LehmannRomano}. Without loss of generality, we take $\E_0\bx<\E_1\bx$, and, for the sake of concreteness, we consider a detector with threshold $\eta$, amounting to the following form for the decision regions:
\beq
\Gamma_0=\{\gamma\in\mathbb{R}:\,
\gamma\leq \eta
\},\qquad \Gamma_1=\mathbb{R}\setminus\Gamma_0.
\label{eq:gamma0_bis}
\eeq
Let us set  $\E_0\bx<\eta<\E_1 \bx$ since, as will be clear soon, choosing a threshold outside the range $(\E_0\bx,\E_1\bx)$ will lead to trivial performance for one of the error exponents. 
According to Theorem~4, to evaluate the exponent ${\cal E}_0$ (resp., ${\cal E}_1$), one must consider the worst-case, i.e., the smallest value of the function $\Omega_0(\gamma)$ (resp., $\Omega_1(\gamma)$), within the corresponding {\em error} region $\Gamma_1$ (resp., $\Gamma_0$). In view of the convexity properties discussed at the end of Sec.~\ref{subsec:mainimplic}, and reported in Appendix C, we see that, for the threshold detector, both minima are attained only at $\gamma=\eta$. 
Certainly, this shape turns out to be of great interest in practical applications where, inspired by the optimality properties of a log-likelihood ratio test in the centralized case, a threshold detector is often an appealing and reasonable choice. 
On the other hand, we would like to stress that different, arbitrary decision regions can be in general chosen, and that the minima of $\Omega_0(\cdot)$ and $\Omega_1(\cdot)$ in Fig.~\ref{fig:Theo4Geo} might be correspondingly located at two different points.

\vspace*{5pt}
In summary, Theorem~4 allows us to compute the exponents ${\cal E}_0$ and ${\cal E}_1$ as functions of $i)$ the kind of statistic $\bx$ employed by the sensors, which determines the shape of the LMGFs $\psi_h(t)$ to be used in~(\ref{eq:Omegah}); and $ii)$ of the employed decision regions relevant for the minimizations in~(\ref{eq:E01}).  
Once ${\cal E}_0$ and ${\cal E}_1$ have been found, the error probabilities $\alpha_{k,\mu}$ and $\beta_{k,\mu}$ can be approximated using Eq.~(\ref{eq:mainres2}). This result is then key for both detector design and analysis, so that we are now ready to illustrate the operation of the adaptive distributed network of detectors.

\section{Examples of Application}
\label{sec:examples}

In this section, we apply the developed theory to four relevant detection problems. We start with the classical Gaussian shift-in-mean problem. 
Then, we consider a scenario of specific relevance for sensor network applications, namely, detection with hardly (one-bit) quantized measurements. This case amounts to testing two Bernoulli distributions with different parameters under the different hypotheses.
Both the Gaussian and the finite-alphabet assumptions are removed in the subsequent example, where a problem of relevance to radar applications is addressed, that is, shift-in-mean with additive noise sampled from a Laplace (double-exponential) distribution. 
Finally, we examine a case where the agents have limited knowledge of the underlying data model, and agree to employ a simple sample-mean detector, in the presence of noise distributed as a Gaussian mixture.
\begin{figure}[t]
\centerline{\boxed{
\includegraphics[width=.25\textheight]{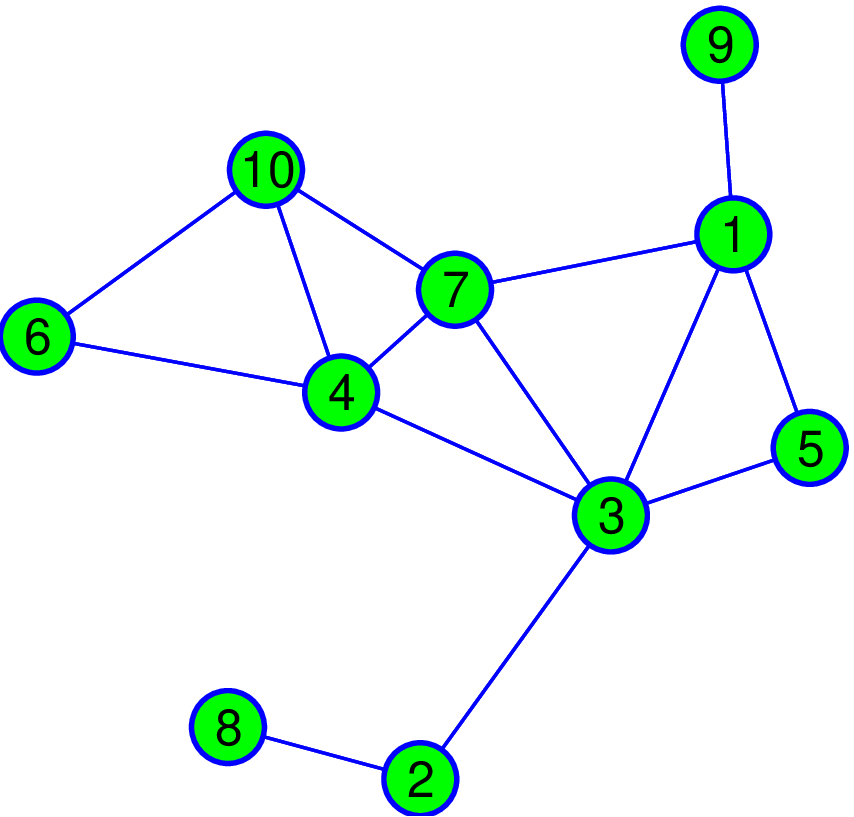}
}
}
\caption{Network skeleton used for the numerical simulations.}
\label{fig:skeleton}
\end{figure}

Before dwelling on the presentation of the numerical experiments, we provide some essential details on the strategy that has been implemented for obtaining them: 
\begin{itemize}
\item
The network used for our experiments consists of ten sensors, arranged so as to form the topology in Fig.~\ref{fig:skeleton}, with  combination weights $a_{k,\ell}$ following the Laplacian rule~\cite{xiao-boyd,SayedAcademicPress}.
\item
The decision rule for the detectors is based on comparing the diffusion output $\by_k(n)$ to some threshold $\eta$, namely,
\beq
\by_k(n)\mathop{\lesseqgtr}_{{\cal H}_1}^{{\cal H}_0} \eta,
\label{eq:TheTest}
\eeq 
where the decision regions are the same as in~(\ref{eq:gamma0_bis}).
\item
Selecting the threshold $\eta$ in~(\ref{eq:TheTest}) is a critical stage of detector design and implementation. 
This choice can be guided by different criteria, which would lead to different threshold settings. In the following examples, we present three relevant cases, namely:  $i)$ a threshold setting that is suited to the Bayesian and the max-min criteria (Sec.~\ref{subsec:bernoulli}); $ii)$ a Neyman-Pearson threshold setting (Sec.~\ref{subsec:Laplace}); $iii)$ and a threshold setting in the presence of insufficient information about the underlying statistical models (Sec.~\ref{subsec:Gmix}).
We would like to stress that using different threshold setting rules for different statistical models has no particular meaning. 
These choices are just meant to illustrate different rules and different models while avoiding repetition of similar results.
\item
The diffusion output is obtained after consultation steps involving the exchange of some local statistics $\bx_k(n)$. 
The particular kind of statistic used in the different examples  will be detailed when needed.



\end{itemize}

\subsection{Shift-in-mean Gaussian Problem}
The first hypothesis testing problem we consider is the following: 
\beqa
{\cal H}_0&:&\bd_k(n)\sim{\cal N}(0,\sigma^2),\\
{\cal H}_1&:&\bd_k(n)\sim{\cal N}(\theta,\sigma^2),
\eeqa
where $\bd_k(n)$ denotes the local datum collected by sensor $k$ at time $n$. 
We assume the local  statistic $\bx_k(n)$ to be shared during the diffusion process is the log-likelihood ratio of the measurement $\bd_k(n)$:
\beq
\bx_k(n)=\frac{\theta}{\sigma^2}\left(
\bd_k(n) - \frac{\theta}{2}
\right).
\eeq
Note that in the Gaussian case the log-likelihood ratio is simply a shifted and scaled version of the collected observation $\bd_k(n)$, such that no substantial differences are expected if the agents share directly the observations.

In the specific case that $\bx_k(n)$ is the log-likelihood ratio, the expectations $\E_0\bx$ and $\E_1\bx$ assume a peculiar meaning. Indeed, they can be conveniently represented as:
\beq
\E_0\bx= - {\cal D}({\cal H}_0||{\cal H}_1),\quad
\E_1\bx={\cal D}({\cal H}_1||{\cal H}_0),
\eeq
where ${\cal D}({\cal H}_i||{\cal H}_j)$, with $i,j\in\{0,1\}$, is the Kullback-Leibler (KL) divergence between hypotheses $i$ and $j$ --- see~\cite{CT}.
In particular, for the Gaussian shift-in-mean problem the distribution of the log-likelihood ratio can be expressed in terms of the KL divergences as follows:
\beq
\bx_k(n)\stackrel{{\cal H}_0}{\sim} {\cal N}(-{\cal D},2{\cal D}),\qquad
\bx_k(n)\stackrel{{\cal H}_1}{\sim} {\cal N}({\cal D},2{\cal D}),
\label{eq:lrdist}
\eeq
where 
\beq
{\cal D}\dfz {\cal D}({\cal H}_0||{\cal H}_1)={\cal D}({\cal H}_1||{\cal H}_0)=\frac{\theta^2}{2\sigma^2},
\eeq 
is the KL divergence for the Gaussian shift-in-mean case~\cite{CT}.

Since the LMGF of a Gaussian random variable ${\cal N}(a,b)$ is $a t +b t^2/2$~\cite{LehmannRomano}, we deduce from~(\ref{eq:lrdist}) that
\beq
\psi_0(t)={\cal D} t (t-1),\quad
\psi_1(t)={\cal D} t (t+1).
\label{eq:LMGFforGauss}
\eeq
Note that $\psi_1(t)=\psi_0(t+1)$, a relationship that holds true more generally when working with the LMGFs of the log-likelihood ratio --- see, e.g.,~\cite{Dembo-Zeitouni}. Now, applying~(\ref{eq:Omegah}) to~(\ref{eq:LMGFforGauss}) readily gives
\beq
\omega_0(t)={\cal D} t\,\left(\frac t 2 -1\right),\quad
\omega_1(t)={\cal D} t\,\left(\frac t 2 +1\right).
\eeq
According to its definition~(\ref{eq:FLtransf}), in order to find the Fenchel-Legendre transform we should maximize, with respect to $t$, the function $\gamma t - \omega(t)$. In view of the convexity properties proved in Appendix C, this can be done by taking the first derivative and equating it to zero, which is equivalent to writing
\beqa
\gamma&=&\omega_0^{\prime}(t_0)=\frac{\psi_0(t_0)}{t_0}\Rightarrow t_0=\frac{\gamma}{{\cal D}}+1,\\
\gamma&=&\omega_1^{\prime}(t_1)=\frac{\psi_1(t_1)}{t_1}\Rightarrow t_1=\frac{\gamma}{{\cal D}}-1.
\eeqa
These expressions lead to
\beq
\Omega_0(\gamma)=\frac{(\gamma+{\cal D})^2}{2{\cal D}},\qquad
\Omega_1(\gamma)=\frac{(\gamma-{\cal D})^2}{2{\cal D}}.
\eeq
Selecting the threshold $\eta$ within the interval $(-{\cal D},{\cal D})$, the minimization in~(\ref{eq:E01}) is easily performed --- refer to Fig.~\ref{fig:Theo4Geo} and the related discussion. The final result is:
\beq
\boxed
{
\alpha_{k,\mu}\stackrel{\cdot}{=}
e^{-(1/\mu)\,S\,\frac{(\eta+{\cal D})^2}{2{\cal D}}},\qquad
\beta_{k,\mu}\stackrel{\cdot}{=}
e^{-(1/\mu)\,S\,\frac{(\eta-{\cal D})^2}{2{\cal D}}}
}
\eeq
These expressions provide the complete asymptotic characterization to the leading exponential order (i.e., they furnish the detection error exponents) of the adaptive distributed network of detectors for the Gaussian shift-in-mean problem, and for any choice of the threshold $\eta$ within the interval $(-{\cal D},{\cal D})$. 

We have run a number of numerical simulations to check the validity of the results. 
Clearly, in order to show the generality of our methods, it is desirable to test them on non-Gaussian data as well. 
Since the interpretation of the results for both Gaussian and non-Gaussian data is essentially similar, we shall skip the numerical results for the Gaussian case to avoid unnecessary repetitions and focus on other cases.
Accordingly, also the discussion on how to make a careful selection of the detection threshold $\eta$ is postponed to the forthcoming sections.

\subsection{Hardly (one-bit) Quantized Measurements}
\label{subsec:bernoulli}

\begin{figure*}
\minipage{0.5\textwidth}
  \vspace*{5pt}
  \includegraphics[width=.35\textheight]{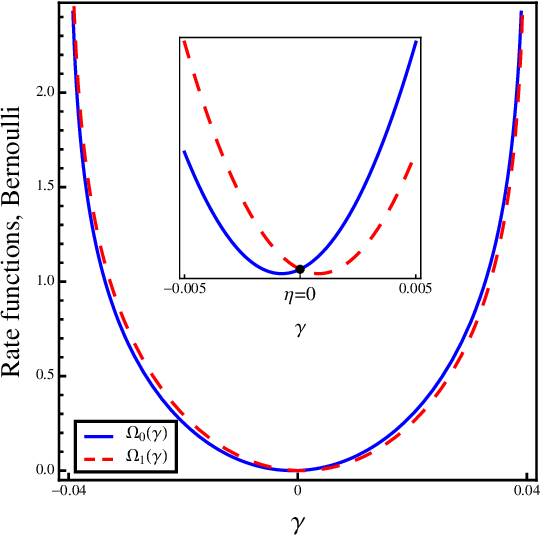}
\endminipage
\minipage{0.5\textwidth}%
 \includegraphics[width=.37\textheight]{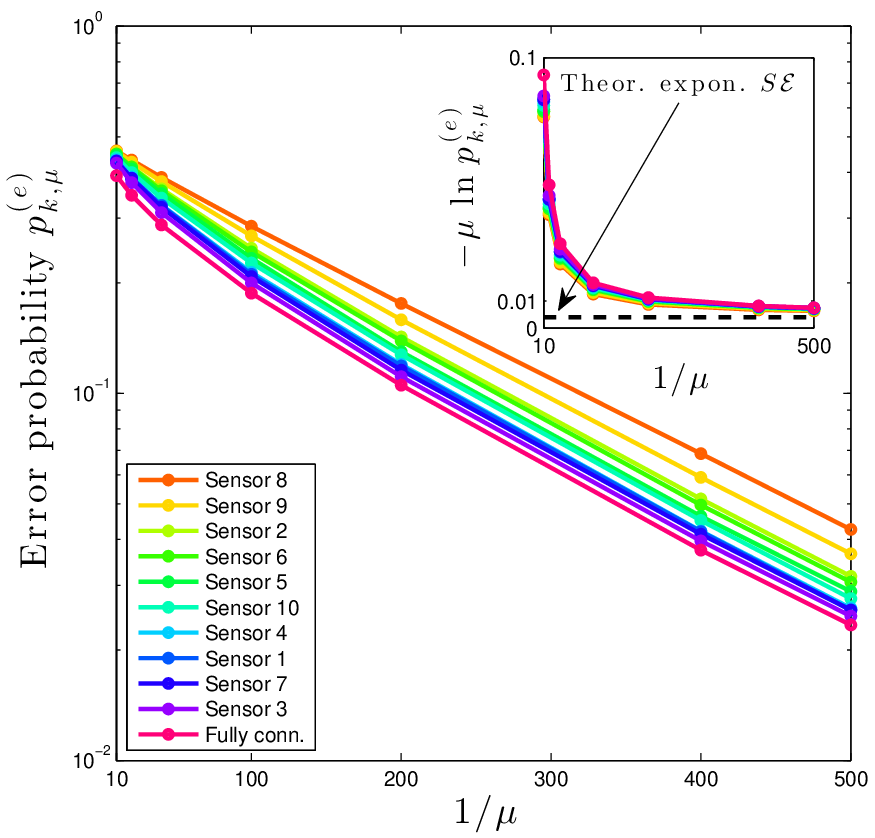}
\endminipage
\caption{Bernoulli example discussed in Sec.~\ref{subsec:bernoulli}. We refer to the network in Fig.~\ref{fig:skeleton}, and use detector~(\ref{eq:TheTest}) with $\eta=0$. 
Leftmost panel: Rate functions. The dark circle in the close-up marks the employed detection threshold, which is relevant to error exponent evaluation.
Rightmost panel: Steady-state error probabilities at different sensors, obtained via Monte Carlo simulation. For comparison purposes, the empirical error probabilities of the fully connected system are reported.
The solid curves in the inset plot represent the empirical error exponents $-\mu\ln p^{(e)}_{k,\mu}$, for $k=1,2,\dots,S$, while the dashed horizontal line is the exponent $S\,{\cal E}$ predicted by our large deviations analysis (Theorem 4). 
The parameters of the considered detection problem are $p_0=0.49$ and $p_1=0.51$. The number of Monte Carlo runs is $10^5$.}
\label{fig:BernoulliTot}
\end{figure*}

We now examine the example in which the measurements at the local sensors are hardly quantized. This situation can be formalized as the following hypothesis test:
\beqa
{\cal H}_0&:&\bd_k(n) \sim{\cal B}(p_0),\label{eq:BernTest0}
\\
{\cal H}_1&:&\bd_k(n) \sim{\cal B}(p_1),
\label{eq:BernTest1}
\eeqa
with ${\cal B}(p)$ denoting a Bernoulli random variable with success probability $p$.
As in the previous example, we assume that the local statistics $\bx_k(n)$ employed by the sensors in the adaptation/combination stages are chosen as the local log-likelihood ratios that, in view of~(\ref{eq:BernTest0})--(\ref{eq:BernTest1}), can be written as:
\beq
\bx_k(n)=\bd_k(n)\ln\left(\frac{p_1}{p_0}\right) +
(1-\bd_k(n))\ln\left(\frac{q_1}{q_0}\right),
\eeq 
where $q_h=1-p_h$, with $h=0,1$.
Since $\bd_k(n)\in\{0,1\}$, we see that $\bx_k(n)$ is a binary random variable taking on the values $\ln(p_1/p_0)$ or $\ln(q_1/q_0)$. The distribution of $\bx_k(n)$ is then characterized by:
\beq
\P_0
\left[\bx_k(n)=\ln\left(
\frac{p_1}{p_0}
\right)
\right]=p_0,
~
\P_1\left[
\bx_k(n)=\ln\left(
\frac{p_1}{p_0}
\right)
\right]=p_1,
\eeq
and, hence, the LMGFs for this example are readily computed:
\beq
\psi_0(t)=\ln\left(\, \frac{p_1^t}{p_0^{t-1}} + \frac{q_1^t}{q_0^{t-1}}\,\right),
\eeq
\beq
\psi_1(t)=\ln\left(\, \frac{p_1^{t+1}}{p_0^t} + \frac{q_1^{t+1}}{q_0^t}\,\right).
\eeq
According to the relationship~(\ref{eq:Omegah}) found in Theorem~4, these closed-form expressions are used for the evaluation of $\omega_0(t)$ and $\omega_1(t)$, which in turn are needed to compute the rate functions $\Omega_0(\gamma)$ and $\Omega_1(\gamma)$. Differently from the Gaussian example, here these tasks need to be performed numerically.
The resulting rate functions are displayed in the leftmost panel of Fig.~\ref{fig:BernoulliTot}, and the observed behavior reproduces what is predicted by the general properties of the rate function --- see also the explanation of Fig.~\ref{fig:RateFunGen}.

Let us now examine the adaptive distributed network of detectors in operation. To do so, we must decide on how to set the detection threshold $\eta$ in~(\ref{eq:TheTest}). 
As a method for selecting the threshold, in this section we illustrate the asymptotic Bayesian criterion that prescribes maximizing the exponent of the average error probability
\beq
p^{(e)}_{k,\mu}\dfz\pi_0 \alpha_{k,\mu} + \pi_1 \beta_{k,\mu},
\eeq 
where $\pi_0$ and $\pi_1$ are the prior probabilities of occurrence of hypotheses ${\cal H}_0$ and ${\cal H}_1$, respectively.
It is easily envisaged that the exponent of the average error probability is determined by the worst one (slowest decay) between the Type-I and Type-II error exponents --- see~\cite[Eq. (I.2), p. 4]{DenHollander}. 
As a result, optimizing the Bayesian error exponent is equivalent to a max-min approach aimed at maximizing the minimum exponent.
We now apply this criterion to the considered example. 
To this aim, a close inspection of the rate functions in Fig.~\ref{fig:BernoulliTot} is beneficial.
First, as it can be seen by the close-up shown in the inset plot, setting the threshold to $\eta=0$ would imply
\beq
{\cal E}_0=\inf_{\gamma>0}\Omega_0(\gamma)=\Omega_0(0)=\Omega_1(0)=
\inf_{\gamma\leq 0}\Omega_1(\gamma)={\cal E}_1\dfz {\cal E}.
\label{eq:01exponents}
\eeq
Moreover, any other choice of the threshold $\eta\neq 0$ makes one of the two exponents smaller than ${\cal E}$. This can be clearly visualized by varying the position of $\eta$ in Fig.~\ref{fig:Theo4Geo}, and computing the infima over the pertinent decision regions. In summary, according to whether we adopt a Bayesian or a max-min criterion, an optimal choice for the threshold in this case is $\eta=0$. 

In the simulations, we refer to a sufficiently large time horizon, such that the steady-state  assumption applies, and evaluate the error probabilities for different values of the step-size --- see the rightmost panel in Fig.~\ref{fig:BernoulliTot}. 
In the considered example, it is easily verified by symmetry arguments that the error probabilities (and not only the exponents) of first and second kind are equal, and therefore they equal the average error probability for any prior distribution of the hypotheses: 
\beq
\alpha_{k,\mu}=\beta_{k,\mu}=p^{(e)}_{k,\mu}.
\eeq
Accordingly, in the following description the terminologies ``error probability" and ``error exponent" can be equivalently and unambiguously referred to any of these errors.

In Fig.~\ref{fig:BernoulliTot}, rightmost panel, the performance of all the agents is displayed as a function of $1/\mu$, and different agents are marked with different colors. For comparison purposes, the performance of the fully connected system is also displayed. 
All these probability curves have been obtained by Monte Carlo simulation.
Some remarkable features are observed.

First, all the different curves pertaining to different agents stay nearly parallel for sufficiently small values of the step-size $\mu$.
This is a way to visualize that $i)$ the detection error probabilities vanish exponentially at rate $1/\mu$; and $ii)$ the detection error {\em exponents} at different sensors are equal, and further equal to that of the fully-connected system corresponding to the centralized stochastic gradient solution.
This is the basic message conveyed by the large deviations analysis. Indeed, the asymptotic relationships for the error probabilities in~(\ref{eq:mainres2})  express convergence {\em to the first leading order in the exponent}.

It remains to show that the {\em exponents} of the simulated error probabilities match the {\em exponents} predicted by Theorem 4.
This is made in the inset plot of Fig.~\ref{fig:BernoulliTot}, rightmost panel, where the horizontal dashed line depicts the theoretical exponent  $S {\cal E}$, with ${\cal E}$ computed using~(\ref{eq:01exponents}), while the solid curves represent the empirical error exponents seen at different sensors, namely the quantities $-\mu\ln p^{(e)}_{k,\mu}$, for $k=1,2,\dots, S$. 
It is observed that, as the step-size decreases, the empirical error exponents converge towards the theoretical one $S\,{\cal E}$. 

A further interesting evidence seems to emerge from the numerical experiments. 
The error probability curves in Fig.~\ref{fig:BernoulliTot}, rightmost panel, are basically ordered. Examining the relationship between this ordering and the sensor placement in Fig.~\ref{fig:skeleton}, it is seen that the ordering reflects the degree of connectivity of each agent. For instance, sensor $3$ has the highest number of neighbors (five), and its performance is the closest to the fully connected case. On the other hand, sensor $8$ is the most isolated, and its error probability curve appears accordingly the highest one.
Note that, since from the presented theory we learned that each agent reaches asymptotically the same detection exponent, these differences are related to higher order corrections (i.e., sub-exponential terms that are neglected in a large deviations analysis) and/or to non-asymptotic effects.
A systematic and thorough analysis of the above features, as well as of their exact interplay with the network connectivity and more in general with the overall structure of the connectivity matrix $A$, requires a refined asymptotic estimate that goes beyond the large deviations analysis carried out here.

\subsection{Shift-in-mean with Laplacian noise}
\label{subsec:Laplace}
In this section we consider another non-Gaussian example, namely, the case of a shift-in-mean detection problem with noise distributed according to a Laplace distribution. Denoting by ${\cal L}(a,b)$ a (shifted) Laplace distribution with shift parameter $a$ and scale parameter $b$, i.e., having the probability density function:
\beq
f_L(\xi)=\frac{1}{2b}e^{-\frac{|\xi-a|}{b}},
\eeq
the hypothesis test we are now interested in is formulated as follows: 
\beqa
{\cal H}_0&:&\bd_k(n)\sim{\cal L}(0,\sigma),\\
{\cal H}_1&:&\bd_k(n)\sim{\cal L}(\theta,\sigma).
\label{eq:LaplaceTest}
\eeqa
We assume again that the local statistics $\bx_k(n)$ are chosen as the local log-likelihood ratios:
\beq
\bx_k(n)=\frac 1 \sigma ( |\bd_k(n)| - |\bd_k(n)-\theta|  ).
\eeq 
Then, the LMGFs for this case can be computed in closed form~\cite{MouraLDnonGauss}, and are given by:
\beqa
\psi_0(t)&=&\ln\left( \frac{1-t}{1-2t}\,e^{-\rho\,t}   -  \frac{t}{1-2t}e^{-\rho\,(1-t)} \right),
\\
\psi_1(t)&=&\ln\left( \frac{1+t}{1+2t}\,e^{\rho\,t}  + \frac{t}{1+2t}e^{-\rho\,(1+t)} \right),
\eeqa
where we defined $\rho=\theta/\sigma$, and where, by limit arguments, we have $\psi_0(1/2)=\psi_1(-1/2)=-\rho/2+\ln(1+\rho/2)$.

\begin{figure*}
\minipage{0.333\textwidth}
  \vspace*{5pt}
  \includegraphics[width=.23\textheight]{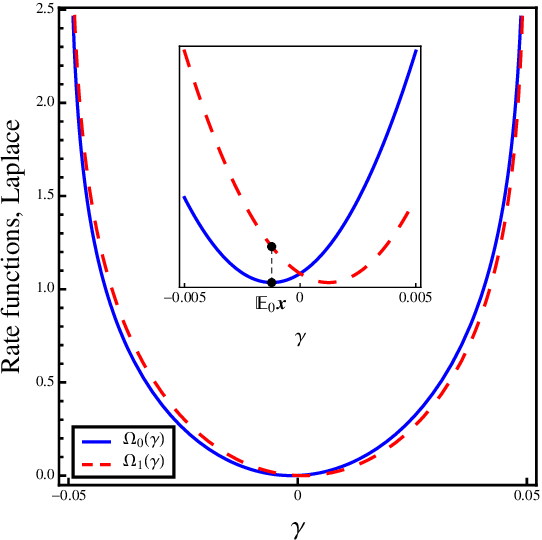}
\endminipage~~
\minipage{0.333\textwidth}%
 \includegraphics[width=.23\textheight]{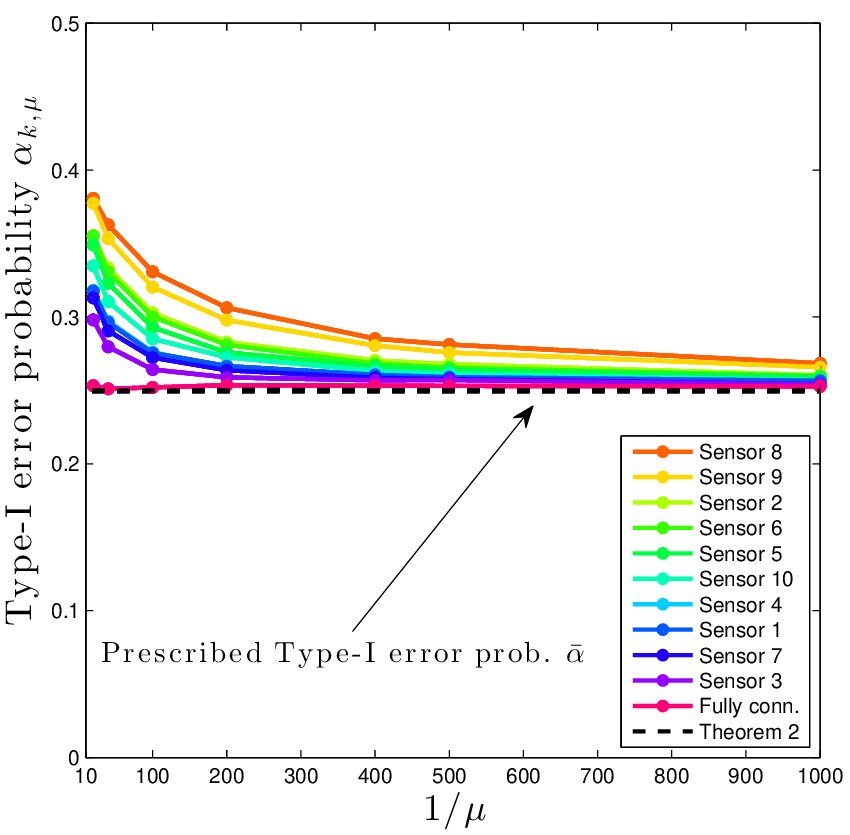}
\endminipage~~
\minipage{0.333\textwidth}%
 \includegraphics[width=.23\textheight]{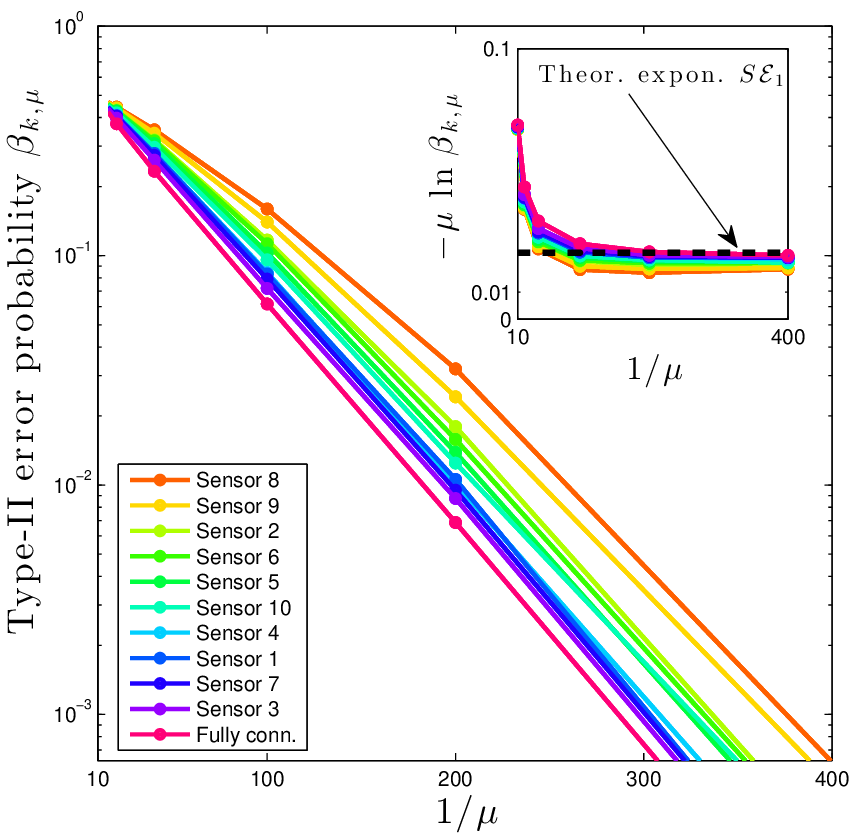}
\endminipage
 \caption{
Laplace example discussed in Sec.~\ref{subsec:Laplace}. 
We refer to the network in Fig.~\ref{fig:skeleton}, and use the Neyman-Pearson detector with threshold~(\ref{eq:NPthresh2}), for two values of the desired false-alarm level $\bar\alpha$.  
Leftmost panel: Rate functions. The dark circle in the close-up marks the abscissa $\eta=\E_0\bx$, which is relevant for computing the Type-II error exponent.  
Middle panel: Solid curves refer to the empirical steady-state Type-I error probabilities at different sensors, obtained via Monte Carlo simulation. For comparison purposes, the empirical error probabilities of the fully connected system are reported. The dashed horizontal lines pertain to the theoretical Type-I error probabilities obtained by the normal approximation (Theorem 2). 
Rightmost panel: Steady-state Type-II  error probabilities at different sensors, along with the performance of the fully connected case.
The solid curves in the inset plot represent the empirical Type-II error exponent $-\mu\ln \beta_{k,\mu}$, for $k=1,2,\dots,S$, while the dashed horizontal line is the exponent predicted by our large deviations analysis (Theorem 4). 
The parameters of the considered detection problem are $\theta=0.05$ and $\sigma=1$. The number of Monte Carlo runs is $10^5$.}
\label{fig:LaplaceTot}
\end{figure*}

As done before, we can use the above expressions in~(\ref{eq:Omegah}), for performing numerical evaluation of $\omega_0(t)$ and $\omega_1(t)$, and of their Fenchel-Legendre transforms $\Omega_0(\gamma)$ and $\Omega_1(\gamma)$, which are displayed in Fig.~\ref{fig:LaplaceTot}, leftmost panel.

Differently from the previous section, we now consider an alternative threshold setting, which is grounded on the well-known Neyman-Pearson criterion~\cite{LehmannRomano}. Its classical (asymptotic) formulation sets a maximum tolerable value for the false-alarm probability, and examines the decaying rate of the miss-detection probability (the role of the two errors can also be reversed).
The main difference in relation to the setup considered so far is that we relax the condition that the Type-I error probability vanish exponentially, and this allows in general for a gain in terms of the Type-II error exponent.
The procedure for the Neyman-Pearson threshold setting has been already described in Sec.~\ref{sec:smallmu} --- see ~(\ref{eq:NPth1})--(\ref{eq:NPpfa1}). Accordingly, to achieve a false-alarm probability $\bar\alpha$, we need a threshold
\beq
\eta=\eta_\mu=\E_0\bx + \sqrt{\frac{\mu \sigma_{x,0}^2}{2 S}}\,Q^{-1}(\bar\alpha).
\label{eq:NPthresh2}
\eeq
It remains to evaluate the Type-II error probability 
\beq
\beta_{k,\mu}=\P_1[\by^\star_{k,\mu}\leq \eta_\mu],
\label{eq:betamuNP}
\eeq
or, more precisely, the corresponding exponent ${\cal E}_1$. For this purpose, we must resort to Theorem 4. Note, however, that the threshold $\eta=\eta_\mu$ now depends on $\mu$ and, hence, Theorem 4 does not directly apply. 
As noted in Remark II, it is instructive to examine how the result of Theorem 4 can be generalized to manage similar situations.
Indeed, we can work in terms of the shifted variables
\beq
\widehat{\by}^\star_{k,\mu}=\by^\star_{k,\mu}-\sqrt{\frac{\mu \sigma_{x,0}^2}{2 S}}\,Q^{-1}(\bar\alpha),
\eeq
yielding
\beq
\beta_{k,\mu}=
\P_1[
\widehat{\by}^\star_{k,\mu}\leq \E_0\bx].
\label{eq:NPexamp}
\eeq
By application of the G\"artner-Ellis Theorem to the shifted variables $\widehat{\by}^\star_{k,\mu}$, it is easy to see that the added  deterministic term (vanishing with $\mu$) does not alter the limiting function $\omega_1(t)$ in~(\ref{eq:Omegah}), and consequently the final rate function~$\Omega_1(\gamma)$. Accordingly, and based on~(\ref{eq:NPexamp}), the Type-II error exponent is
\beq
{\cal E}_1=\inf_{\gamma\leq\E_0\bx}\Omega_1(\gamma)=\Omega_1(\E_0\bx).
\label{eq:NPexpo}
\eeq
The main implication of the above result can be understood, e.g., by examining the close-up in the leftmost panel of Fig.~\ref{fig:LaplaceTot}, where it is seen that:
\beq
{\cal E}_1=\Omega_1(\E_0\bx)>
\Omega_1(0),
\eeq
the latter value being the Type-II error exponent achieved by the max-min optimal detector with zero threshold previously described.
This immediately shows the gain achieved by relaxing the constraint that {\em both} error probabilities must vanish exponentially.

We now present the numerical evidence for the Neyman-Pearson adaptive distributed detector.
The middle panel in Fig.~\ref{fig:LaplaceTot} shows the convergence of $\alpha_{k,\mu}$ towards the prescribed Type-I error probability $\bar\alpha$ as the step-size $\mu$ goes to zero.
The rightmost panel refers instead to the corresponding Type-II error probability curves. The conclusions that can be drawn are similar to those discussed in the previous example, confirming the validity of the theoretical analysis. It is also interesting to note that the ordering of the different curves, for both error probabilities, is exactly the same obtained in the Bernoulli example. Since the network employed for the simulations is unchanged, this is another clue that the ordering may be related to the structure of the connectivity matrix $A$.

\subsection{Shift-in-mean with Gaussian mixture noise}
\label{subsec:Gmix}

\begin{figure*}
\minipage{0.333\textwidth}
  \vspace*{5pt}
  \includegraphics[width=.22\textheight]{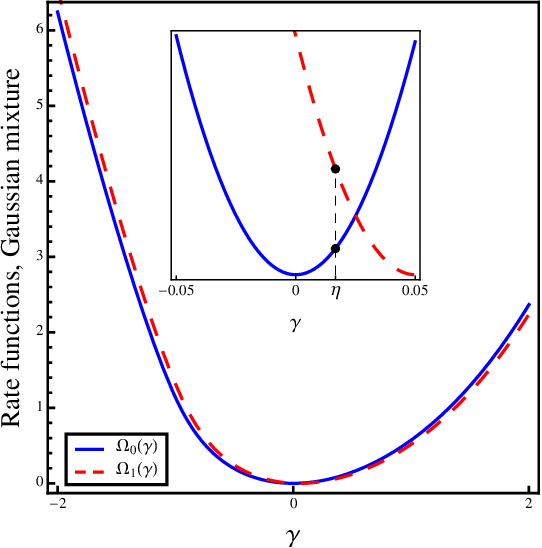}
\endminipage~~
\minipage{0.333\textwidth}%
 \includegraphics[width=.23\textheight]{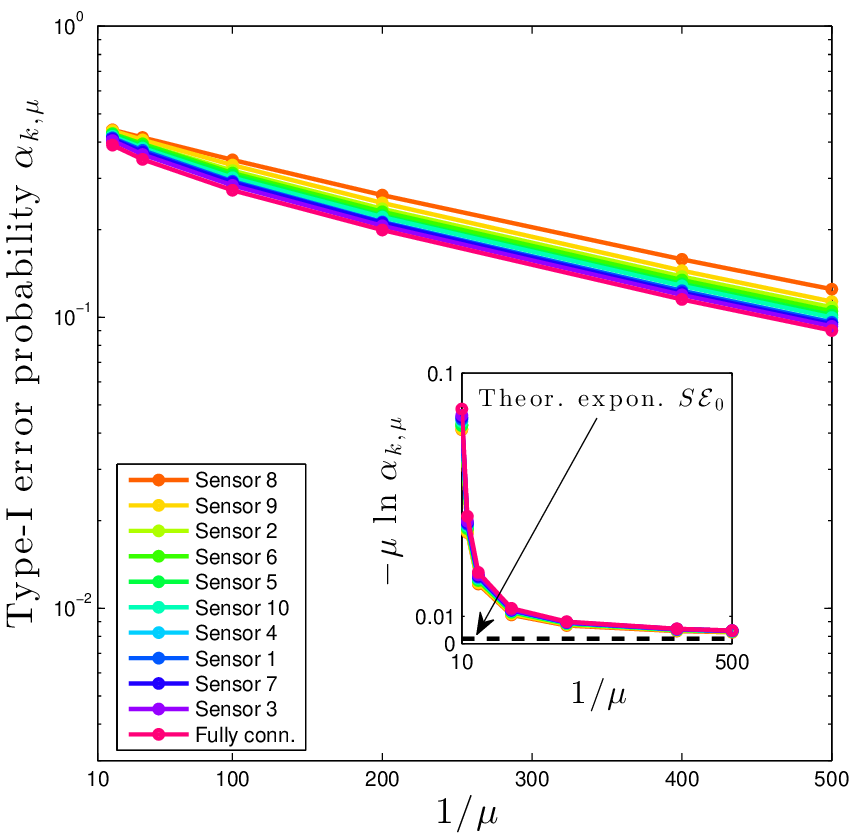}
\endminipage~~
\minipage{0.333\textwidth}%
 \includegraphics[width=.23\textheight]{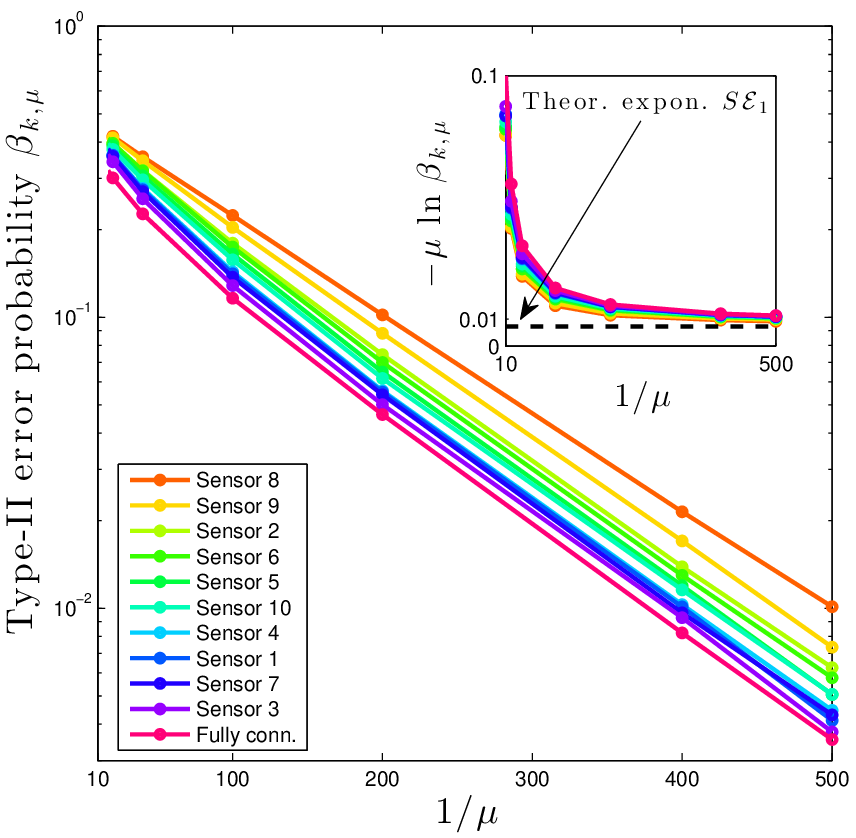}
\endminipage
 \caption{
Gaussian mixture example discussed in Sec.~\ref{subsec:Gmix}. 
We refer to the network in Fig.~\ref{fig:skeleton}, 
and use detector~(\ref{eq:TheTest}) with $\eta=\theta/3$. 
Leftmost panel: Rate functions. The dark circle in the close-up marks the employed detection threshold, which is relevant for evaluating the error exponents.
Middle panel: 
Steady-state Type-I  error probabilities at different sensors, obtained via Monte Carlo simulation. For comparison purposes, the empirical error probabilities of the fully connected system are reported. 
The solid curves in the inset plot represent the empirical Type-I error exponent $-\mu\ln \alpha_{k,\mu}$, for $k=1,2,\dots,S$, while the dashed horizontal line is the exponent predicted by our large deviations analysis (Theorem 4). 
Rightmost panel: Same of middle panel, but for the Type-II error.
The parameters of the considered detection problem are $\theta=0.05$, $\theta_0=1$, $\sigma_1=1$, and $\sigma_2=0.3$. The number of Monte Carlo runs is $10^5$.}
\label{fig:GmixTot}
\end{figure*}

As a final example, we consider the case of a shift-in-mean detection problem with noise distributed according to a zero-mean Gaussian mixture, having the probability density function
\beq
f_{GM}(\xi)=\frac{1}{2} \left(
\frac{1}{\sqrt{2\pi b_1}}e^{-\frac{(\xi-a_0)^2}{2 b_1}}
+
\frac{1}{\sqrt{2\pi b_2}}e^{-\frac{(\xi+a_0)^2}{2 b_2}}
\right),
\eeq
namely, a balanced mixture of normal random variables with different variances $b_1$ and $b_2$, and symmetric expectations $\pm a_0$.
Denoting by ${\cal N}_{mix}(a,a_0,b_1,b_2)$ a {\em shifted} Gaussian mixture distribution with shift parameter $a$, we consider the following hypothesis test:
\beqa
{\cal H}_0&:&\bd_k(n) \sim{\cal N}_{mix}(0,\theta_0,\sigma^2_1,\sigma^2_2),\\
{\cal H}_1&:&\bd_k(n) \sim{\cal N}_{mix}(\theta,\theta_0,\sigma^2_1,\sigma^2_2).
\label{eq:GMTest}
\eeqa
For this model, we do {\em not} assume that the local statistics $\bx_k(n)$ are chosen as the local log-likelihood ratios. 
We assume instead that the agents of the network have scarce knowledge about the underlying statistical model. They know that it is a shift-in-mean problem, and possess a rough information about the value of $\theta$.  
In these circumstances, the agents decide to implement a distributed sample-mean detector, namely, they exchange the local measurements {\em as they are, without any additional pre-processing}. This amounts to state that
\beqa
{\cal H}_0&:&\bx_k(n)\sim{\cal N}_{mix}(0,\theta_0,\sigma^2_1,\sigma^2_2),\\
{\cal H}_1&:&\bx_k(n)\sim{\cal N}_{mix}(\theta,\theta_0,\sigma^2_1,\sigma^2_2).
\label{eq:GMTest}
\eeqa
Then, the LMGFs for this case can be computed in closed form~\cite{MouraLDnonGauss}, and are given by:
\beqa
\psi_0(t)&=&\ln\left( \frac 1 2 e^{\theta_0 t +\frac{\sigma_1^2 t^2}{2}} + \frac 1 2 e^{-\theta_0 t +\frac{\sigma_2^2 t^2}{2}}\right),
\\
\psi_1(t)&=&\theta t +\psi_0(t).
\eeqa
The above expressions are used in~(\ref{eq:Omegah}) for evaluating numerically $\omega_0(t)$ and $\omega_1(t)$, and then their Fenchel-Legendre transforms $\Omega_0(\gamma)$ and $\Omega_1(\gamma)$. 
These latter are depicted in the leftmost panel of Fig.~\ref{fig:GmixTot}. 
We assume the agents in the network are not able to optimize the choice of the detection threshold, due to their limited knowledge of the underlying statistical models. The particular value used in the simulations is $\eta=\theta/3$, which is marked in the close-up of Fig.~\ref{fig:GmixTot}, leftmost panel. It is seen that, differently from the previous examples, this choice does not correspond to a balancing of the detection error exponents, such that it is expected that the Type-I and Type-II error probabilities behave quite differently in this case. This is clearly observed in the middle (Type-I error) and rightmost (Type-II error) panels of Fig.~\ref{fig:GmixTot}.
The numerical evidence confirms the theoretical predictions, as well as the essential features found in all the previous examples. Moreover, it is seen that the enhanced decaying rate of the Type-II error probability arising from the unbalanced threshold setting is paid in terms of a higher Type-I error probability.

\subsection{Adaptation and detection}
In the simulation results illustrated so far, we focused on the system performance at steady-state. 
It is of great interest to consider also the {\em time-evolution} of the system performance, and even more to show the system at work in a {\em dynamic} situation where the true hypothesis is changing over time, which is truly the main motivation for an {\em adaptive} framework. 

\begin{figure}[t]
\centerline{\includegraphics[width=.55\textheight]{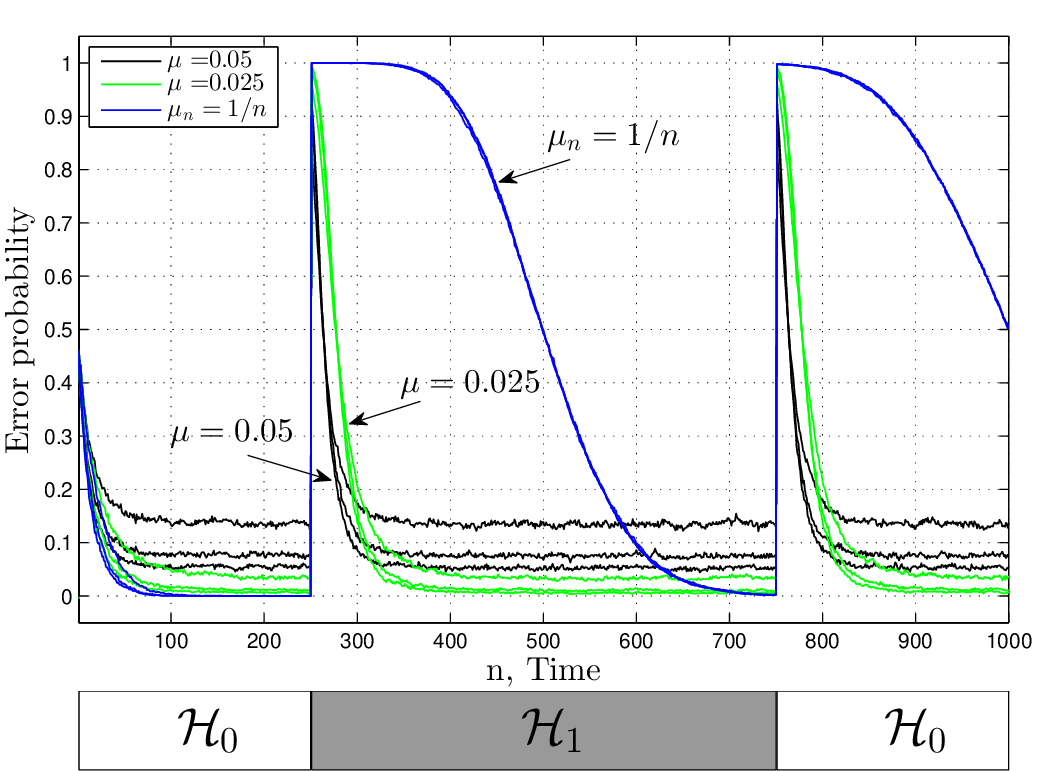}}
\caption{Pictorial summary of adaptive diffusion for detection, with reference to the Laplace example discussed in Sec.~\ref{subsec:Laplace}. Top panel: time-evolution of the error probability at a local node with $i)$ the diffusion strategy with different step-sizes $\mu=0.025, 0.05$, and $ii)$ the running consensus strategy (diminishing step-size $\mu_n=1/n$). Actual variation of the true hypothesis is depicted in the bottom panel. The parameters of the considered detection problem are $\theta=0.3$ and $\sigma=1$.}
\label{fig:finalerrprob}
\end{figure}

To this aim, we return to the kind of situation described in Fig.~\ref{fig:adaptection}, which is now re-examined in more quantitative terms by focusing on the actual error probabilities, rather than on the time-evolution of the detection statistics. 
Specifically, in Fig.~\ref{fig:finalerrprob} we display the performance of three generic agents of the network, for two values of the step-size. For comparison purposes, we show also the performance of the running consensus algorithm~\cite{running-cons,asymptotic-rc,Bracaetal-Pageconsensus,MouraLDGauss,MouraLDnonGauss,MouraLDnoisy}. 
The underlying statistical model is the shift-in-mean with Laplacian noise detailed in Sec.~\ref{subsec:Laplace}, and we employ a zero-threshold detector. 

First, the inference/adaptation trade-off is emphasized: smaller values of $\mu$ allow better inference (lower values of the steady-state error probabilities), at the cost of increasing the time for reliably learning that a change occurred.
In this respect, the running consensus performance represents an  extreme case: indeed, here the step-size is vanishing, i.e., $\mu_n=1/n$, which explains the bad performance in terms of adaptation exhibited in Fig.~\ref{fig:adaptection}.

\section{Concluding Remarks and Open Issues}
\label{sec:conclu}
The asymptotic tools developed in this paper allow designing and characterizing the performance of network detectors that are {\em adaptive and decentralized}. 
We show that the steady-state detection error probabilities of each individual agent decrease exponentially with the inverse of the step-size and that cooperation among sensors makes the error exponents governing such decay equal to that of a centralized stochastic gradient solution.
Closed-form expressions are derived, giving insights about the main scaling laws with respect to the fundamental system parameters.

In our treatment, we studied the detection performance of the diffusion strategy, {\em given a certain local statistic} $\bx$. 
Our findings show that the steady-state observable, as well as its detection performance, in general depend upon the kind of transmitted data $\bx$. A plausible, though heuristic, choice for $\bx$ is that of the log-likelihood ratio of the measured data. However, the problem of choosing the {\em best} statistic is open, and we feel that the obtained results can assist in exploring the relationship between the asymptotic performance and the choice of an optimal statistic $\bx$.

We would like to finally note that in order to avoid a prohibitive number of Monte-Carlo runs, the simulations in the previous section were run in the small signal-to-noise ratio regime, where the error probabilities need not be too small. In this regime, the exact rate functions could in principle be replaced by parabolic approximations (see, e.g., the leftmost plot in Fig.~\ref{fig:BernoulliTot}) and a parabolic approximation is basically a Gaussian approximation. To avoid confusion, we note that the results of this work  do not require any small signal-to-noise ratio assumption; they hold in greater generality. Moreover, using a Gaussian approximation will generally lead to a wrong error exponent. For the same reason of avoiding prohibitive simulation runs in the convergence analysis of the Type-II error exponent, the Type-I error probability for the Neyman-Pearson setting of Fig.~\ref{fig:LaplaceTot} was set to $\bar\alpha=1/4$ (rather than to much smaller values) and used to illustrate the theoretical findings against the simulated curves.

\section*{Appendix A: Proof of Theorem 2}
Since the transient term in~(\ref{eq:mainATC}) does not affect the limiting behavior of $\by_k(n)$, it suffices to focus on the limiting behavior of the summations in~(\ref{eq:sumexpress}). We introduce accordingly the following finite-horizon variable:
\beq
\by_k^{\star}(n)\dfz \sum_{i=1}^n \bz_k(i)\;=\;
\sum_{i=1}^n \sum_{\ell=1}^S \mu (1-\mu)^{i-1} b_{k,\ell}(i)\bx_{\ell}^{\prime}(i).
\label{eq:regimeterm}
\eeq
Since  $\by^{\star}_k(n)$ converges in distribution to $\by^\star_{k,\mu}$ as $n\rightarrow\infty$, by L\'evy's continuity Theorem~\cite{FellerBookV2},  the corresponding characteristic functions must converge as well. 
It is convenient to work in terms of the normalized variable:
\beq
\widetilde\by^\star_{k,\mu}=\frac{\by^\star_{k,\mu}-\E\bx}{\sqrt{\mu\,\sigma_x^2/(2S)}}.
\label{eq:centeredy}
\eeq
Denoting by $\varphi_{k,\mu}(t)$ the characteristic function of $\widetilde \by^\star_{k,\mu}$, using~ (\ref{eq:regimeterm}) and~(\ref{eq:centeredy}), and taking the limit as $n\rightarrow\infty$, we have:
\beq
\varphi_{k,\mu}(t)=\E e^{j t \widetilde\by^\star_{k,\mu}}
=
\prod_{i=1}^{\infty}\prod_{\ell=1}^S \E e^{j t \widetilde\bx^{\prime}_{\ell}(i) \zeta_{i,\ell}},
\eeq
defined in terms of the non-random variable 
\beq
\zeta_{i,\ell}=\sqrt{2 S\mu}(1-\mu)^{i-1}b_{k,\ell}(i),
\label{eq:zetadefin}
\eeq 
and the centered and normalized random variable 
\beq
\widetilde{\bx}_{\ell}^{\prime}(i)=\frac{\bx_{\ell}^{\prime}(i)-\E\bx}{\sigma_x}.
\eeq
Now, the claim of asymptotic normality in~(\ref{eq:CLT}) can be proven by showing the convergence, as $\mu\rightarrow 0$, of $\varphi_{k,\mu}(t)$ towards the characteristic function of the standard normal distribution, $e^{-\frac{t^2}{2}}$. 
It suffices to work with $t>0$ to verify the validity of this latter property.
Formally, we would like to show that the quantity: 
\beq
\left|\varphi_{k,\mu}(t)-e^{-\frac{t^2}{2}}\right|
=\left|\prod_{i=1}^{\infty}\prod_{\ell=1}^S \E e^{j t \widetilde\bx_{\ell}^{\prime}(i) \zeta_{i,\ell}}-e^{-\frac{t^2}{2}}\right|
\label{eq:claimclaim}
\eeq
converges to zero as $\mu\rightarrow 0$.
To this aim, we start by working with a finite $n$, and write:
\beqa
\lefteqn{
\hspace*{-20pt}
\left|\prod_{i=1}^{n}\prod_{\ell=1}^S \E e^{j t \widetilde\bx_{\ell}^{\prime}(i) \zeta_{i,\ell}} - e^{-\frac{t^2}{2}}\right|
}
\nonumber\\
&\leq&
\left|\prod_{i=1}^{n}\prod_{\ell=1}^S \E e^{j t \widetilde\bx_{\ell}^{\prime}(i) \zeta_{i,\ell}} - 
\prod_{i=1}^{n}\prod_{\ell=1}^S e^{-\frac{t^2 \zeta_{i,\ell}^2}{2}}
\right|\nonumber\\
&+&
\left|
\prod_{i=1}^{n}\prod_{\ell=1}^S e^{-\frac{t^2 \zeta_{i,\ell}^2}{2}}
 - e^{-\frac{t^2}{2}}\right|.
\label{eq:boundbound}
\eeqa
We first focus on the first term on the RHS of~(\ref{eq:boundbound}).
For complex $w_i,z_i$, with $|w_i|\leq 1$ and $|z_i|\leq 1$, it is known that~\cite{FellerBookV2}: 
\beq
\left|\prod_{i=1}^n w_i-\prod_{i=1}^n z_i\right|
\leq\sum_{i=1}^n |w_i-z_i|.
\label{eq:prodsumineq}
\eeq 
Since $\E e^{j t \widetilde\bx_{\ell}^{\prime}(i) \zeta_{i,\ell}}$ is a characteristic function, its magnitude is not greater than one~\cite{FellerBookV2}, such that it is legitimate to write, in view of~(\ref{eq:prodsumineq}):
\beqa
\lefteqn{
\hspace*{-40pt}
\left|\prod_{i=1}^{n}\prod_{\ell=1}^S \E e^{j t \widetilde\bx_{\ell}^{\prime}(i) \zeta_{i,\ell}}
-
\prod_{i=1}^{n}\prod_{\ell=1}^S e^{-\frac{t^2 \zeta_{i,\ell}^2}{2}}\right|
}
\nonumber\\
&&\leq
\sum_{i=1}^{n}\sum_{\ell=1}^S
\left|
\E e^{j t \widetilde\bx_{\ell}^{\prime}(i) \zeta_{i,\ell}}
-
e^{-\frac{t^2 \zeta_{i,\ell}^2}{2}}
\right|.
\eeqa
The single summand on the right-hand side of the above expression is upper bounded by
\beq
\left|
\E e^{j t \widetilde\bx_{\ell}^{\prime}(i) \zeta_{i,\ell}}
-
1+\frac{t^2\zeta^2_{i,\ell}}{2}
\right|
+
\left|
e^{-\frac{t^2 \zeta_{i,\ell}^2}{2}}
-1+\frac{t^2\zeta^2_{i,\ell}}{2}\right|.
\label{eq:twoterms}
\eeq
Using the fact that $\E\widetilde{\bx}_{\ell}^{\prime}(i)=0$ and $\E[\widetilde{\bx}_{\ell}^{\prime}(i)]^2=1$, we can further bound the first term in the above expression as 
\beqa
\lefteqn{\hspace*{-5pt}
\left|
\E e^{j t \widetilde\bx_{\ell}^{\prime}(i) \zeta_{i,\ell}}
-
1+\frac{t^2\zeta^2_{i,\ell}}{2}
\right|
}
\nonumber\\
&=&
\left|
\E \left(
e^{j t \widetilde\bx_{\ell}^{\prime}(i) \zeta_{i,\ell}}
-
1
-
j\widetilde\bx_{\ell}^{\prime}(i) t \zeta_{i,\ell}
+
[\widetilde\bx_{\ell}^{\prime}(i)]^2\,\frac{t^2\zeta^2_{i,\ell}}{2}
\right)
\right|\nonumber\\
&\leq&
\E|\widetilde\bx_{\ell}^{\prime}(i)|^3\,
\frac{t^3 \zeta_{i,\ell}^3}{6},
\eeqa
where the last inequality follows from upper bounding the remainder of the Taylor expansion of the complex exponential:
\beq
\left|
e^{j t} -1 -\frac{jt}{1!}-\dots-\frac{(j t)^{n-1}}{(n-1)!}
\right|
\leq
\frac{|t|^n}{n!}.
\eeq 
Likewise, the second term in~(\ref{eq:twoterms}) is upper bounded by $\frac{t^4\zeta^4_{i,\ell}}{8}$ since $|e^{-s}-1+s|\leq s^2/2$ for any $s\geq 0$.

We can accordingly rewrite~(\ref{eq:boundbound}) as:
\beqa
\lefteqn{
\hspace*{-20pt}
\left|\prod_{i=1}^{n}\prod_{\ell=1}^S \E e^{j t \widetilde\bx_{\ell}^{\prime}(i) \zeta_{i,\ell}} - e^{-\frac{t^2}{2}}\right|
}\nonumber\\
&\leq&
\E|\widetilde\bx_{\ell}^{\prime}(i)|^3\,
\frac{t^3}{6}\,
\sum_{i=1}^{n}\sum_{\ell=1}^S
\zeta_{i,\ell}^3
\nonumber\\
&+&
\frac{t^4}{8}\,
\sum_{i=1}^{n}\sum_{\ell=1}^S
\zeta_{i,\ell}^4\nonumber\\
&+&
\left|e^{-\frac{t^2}{2} 
\sum_{i=1}^{n}\sum_{\ell=1}^S
\zeta^2_{i,\ell}} - e^{-\frac{t^2}{2}}\right|.
\nonumber\\
\label{eq:takethelimits}
\eeqa
We now take the limit as $n\rightarrow\infty$ in the above expression. To this aim, observe that, by the definition~(\ref{eq:zetadefin}), the summation:
\beq
\sum_{i=1}^{n}\sum_{\ell=1}^S\zeta^{m}_{i,\ell}, \qquad m=1,2,\dots
\label{eq:serieszetaterms}
\eeq
is made of nonnegative terms, and is upper bounded by a convergent geometric series, since $b_{k,\ell}\leq 1$.
This implies the convergence of the series~(\ref{eq:serieszetaterms}) as $n\rightarrow\infty$.
Accordingly, taking the limit as $n\rightarrow\infty$ in~(\ref{eq:takethelimits}), and using~(\ref{eq:claimclaim}), we have:
\beqa
\left|\varphi_{k,\mu}(t)-e^{-\frac{t^2}{2}}\right|&=&
\left|\prod_{i=1}^{\infty}\prod_{\ell=1}^S \E e^{j t \widetilde\bx_{\ell}^{\prime}(i) \zeta_{i,\ell}}-e^{-\frac{t^2}{2}}\right|
\nonumber\\
&\leq&
\E|\widetilde\bx_{\ell}^{\prime}(i)|^3\,
\frac{t^3}{6}\,
\sum_{i=1}^{\infty}\sum_{\ell=1}^S
\zeta_{i,\ell}^3
\nonumber\\
&+&
\frac{t^4}{8}\,
\sum_{i=1}^{\infty}\sum_{\ell=1}^S
\zeta_{i,\ell}^4\nonumber\\
&+&
\left|e^{-\frac{t^2}{2} 
\sum_{i=1}^{\infty}\sum_{\ell=1}^S
\zeta^2_{i,\ell}} - e^{-\frac{t^2}{2}}\right|.
\nonumber\\
\eeqa
According to the latter relationships, in order to show that $\left|\varphi_{k,\mu}(t)-e^{-\frac{t^2}{2}}\right|$ converges to zero as $\mu\rightarrow 0$, it suffices to verify that:
\beqa
&&\sum_{i=1}^{\infty}\sum_{\ell=1}^S\zeta^{m}_{i,\ell}\stackrel{\mu\rightarrow 0}{\longrightarrow} 0,
\qquad m=3,4,
\label{eq:terms34}
\\
&&\sum_{i=1}^{\infty}\sum_{\ell=1}^S\zeta^2_{i,\ell}\stackrel{\mu\rightarrow 0}{\longrightarrow} 1.
\label{eq:term2}
\eeqa
A technical remark is useful at this stage. Given the assumption of finite absolute third moment, there exists a simpler way to prove our claim, relying on the celebrated Berry-Esseen theorems~\cite[p. 542]{FellerBookV2}. Such technique would directly reduce our proof  to the verification of properties such as~(\ref{eq:terms34}) and~(\ref{eq:term2}), without the preliminary work with characteristic functions. However, we prefer to offer here a more general proof, which might be useful to obtain future generalizations where the condition about the third moment could be weakened.

The key for proving~(\ref{eq:terms34}) and~(\ref{eq:term2}) is Perron's Theorem, which provides a uniform bound on the convergence rate of the matrix $B_n=A^n$ --- see~\cite[Th. 8.5.1]{Johnson-Horn}.
Let $\lambda_2$ be the second largest magnitude eigenvalue of $A$. For any positive $\lambda$ such that $|\lambda_2|<\lambda<1$, there exists a positive constant ${\cal C}={\cal C}(\lambda,A)$, ensuring for all $i, k$ and $\ell$:
\beq
\left |b_{k,\ell}(i)-\frac 1 S\right |\leq {\cal C}\lambda^i.
\label{eq:Perron}
\eeq
The above result follows by noting that the largest magnitude eigenvalue of the difference matrix $B_n-(1/S)\,\mathds{1}\mathds{1}^T$ is $\lambda_2$, and by applying the result on the convergence rate in~\cite[Corollary 5.6.13]{Johnson-Horn}.

According to the above discussion, let us modify the variables $\zeta_{i,\ell}$ by replacing the matrix entries $b_{k,\ell}(i)$ with their limit $1/S$, namely,
\beq
\widetilde \zeta_{i,\ell}=\sqrt{2 S\mu}(1-\mu)^{i-1}\frac 1 S,
\eeq 
and introduce, for any integer $m\geq 2$, the absolute difference:
\beqa
\lefteqn{
\left|
\sum_{i=1}^{\infty}\sum_{\ell=1}^S
\zeta^m_{i,\ell}
-
\sum_{i=1}^{\infty}\sum_{\ell=1}^S
\widetilde\zeta^m_{i,\ell}
\right|
\leq
\sum_{i=1}^{\infty}\sum_{\ell=1}^S
|\zeta^m_{i,\ell}-\widetilde\zeta^m_{i,\ell}|
}
\nonumber\\
&=&(2S\mu)^{m/2}
\,\sum_{i=1}^{\infty}\sum_{\ell=1}^S
(1-\mu)^{m(i-1)}\,\left|b^m_{k,\ell}(i)-\frac{1}{S^m}\right|.\nonumber\\
\eeqa
Recalling the factorization
\beq
a^m-b^m=(a-b)\sum_{k=0}^{m-1} a^k b^{m-1-k},
\eeq
(which can be proved, for $a\neq b$, by using the geometric sum $\sum_{k=0}^{m-1} r^k=\frac{1-r^m}{1-r}$, and using $r=a/b$), along with the fact that $b_{k,\ell}(i)$ and $1/S$ are not greater than one, we conclude that
\beq
\left|b^m_{k,\ell}(i)-\frac{1}{S^m}\right|\leq
 m \left|b_{k,\ell}(i)-\frac{1}{S}\right|,
\eeq
yielding
\beqa
\lefteqn{
\left|
\sum_{i=1}^{\infty}\sum_{\ell=1}^S
\zeta^m_{i,\ell}
-
\sum_{i=1}^{\infty}\sum_{\ell=1}^S
\widetilde\zeta^m_{i,\ell}
\right|
}\nonumber\\
&\leq&
m (2S\mu)^{m/2}
\,\sum_{i=1}^{\infty}\sum_{\ell=1}^S
(1-\mu)^{m(i-1)}\,\left|b_{k,\ell}(i)-\frac{1}{S}\right|
\nonumber\\
&\leq&
{\cal C}\lambda m
(2S\mu)^{m/2}
\sum_{i=1}^{\infty}\sum_{\ell=1}^S
(1-\mu)^{m(i-1)} \lambda^{i-1}\nonumber\\
&=&
{\cal C}\lambda m
2^{m/2}S^{m/2+1}\,
\frac{\mu^{m/2}}{1-\lambda (1-\mu)^m}\stackrel{\mu\rightarrow 0}{\longrightarrow}0,
\eeqa
where the second inequality follows from~(\ref{eq:Perron}), and the limit holds because $\lambda<1$. In view of the above result, 
in order to establish~(\ref{eq:terms34}) and~(\ref{eq:term2})  it is enough to study the limiting behavior of the summation:
\beqa
\sum_{i=1}^{\infty}\sum_{\ell=1}^S
\widetilde\zeta^m_{i,\ell}&=&
\frac{(2\mu)^{m/2}}{S^{m/2-1}}
\sum_{i=1}^{\infty}
(1-\mu)^{m(i-1)}\nonumber\\
&=&
\frac{2^{m/2}}{S^{m/2-1}}\,
\frac{\mu^{m/2}}{1-(1-\mu)^m}.
\eeqa
Applying L'Hospital's rule~\cite{RudinBook}, the limit of the RHS as $\mu\rightarrow 0$ is: 
\beq
\left(\frac{2}{S}\right)^{m/2-1}\,\lim_{\mu\rightarrow 0} \frac{\mu^{m/2-1}}{(1-\mu)^{m-1} },
\label{eq:DelHopital}
\eeq
which converges to $1$ for $m=2$, and to $0$ otherwise, completing the proof.

 
\section*{Appendix B: Proof of Theorem 3}
We first list some regularity properties of $\psi(t)$ that will be applied in the subsequent analysis --- see, e.g.,~\cite{Dembo-Zeitouni,DenHollander}: 
\begin{enumerate}
\item
By assumption, $\psi(t)<\infty$ for all $t\in\mathbb{R}$. Since it is a LMGF, it is infinitely differentiable in $\mathbb{R}$.
Also, since $\bx$ is a non-degenerate (i.e., non deterministic) random variable, we have
\beq
\psi^{\prime\prime}(t)>0,\qquad \forall t\in\mathbb{R},
\label{eq:psi2der}
\eeq
and, hence, $\psi(t)$ is strictly convex in $\mathbb{R}$.
\item
With reference to the function $\frac{\psi(t)}{t}$ appearing in~(\ref{eq:omegadef}), we note that
\beq
\lim_{t\rightarrow 0} \frac{\psi(t)}{t}=\psi^\prime(0),
\label{eq:psiprime0}
\eeq 
and, hence, $\frac{\psi(t)}{t}$ is continuous for all $t\in\mathbb{R}$, and the integral in~(\ref{eq:omegadef}) is well-posed. 
\item
For all $t\neq 0$, we have
\beq
\frac{d}{dt}\frac{\psi(t)}{t}=\frac{\psi^\prime(t)\,t-\psi(t)}{t^2},
\label{eq:dertneq0}
\eeq
with 
\beq
\lim_{t\rightarrow 0}\frac{\psi^\prime(t)\,t-\psi(t)}{t^2}=\frac{\psi^{\prime\prime}(0)}{2},
\label{eq:tequal0}
\eeq
implying that $\frac{d}{dt}\frac{\psi(t)}{t}$ is continuous for all $t\in\mathbb{R}$. 
In addition, we have:
\beq
\frac{d}{dt} \frac{\psi(t)}{t}>0,\quad \forall t\in\mathbb{R}.
\label{eq:posder}
\eeq
This is immediately verified for $t=0$ by using~(\ref{eq:psi2der}) in~(\ref{eq:tequal0}). 
For $t\neq0$, since $\psi(t)$ is strictly convex and differentiable in $\mathbb{R}$, we can apply the first-order condition for strict convexity --- see Eq.~(3.3) in~\cite{boyd-vandenberghe}:
\beq
\psi(a)-\psi(b)>\psi^{\prime}(b)(a-b),\quad \forall a,b\in\mathbb{R},\quad a\neq b.
\label{eq:firstordcond}
\eeq
Setting $a=0$, $b=t\neq 0$, and using $\psi(0)=0$, result~(\ref{eq:posder}) now follows from~(\ref{eq:dertneq0}).

\end{enumerate}

\vspace*{10pt}
\noindent
In the following, we denote by $\phi^{(c)}_{\mu}(t)$ the LMGF of the steady-state variable $\by_{k,\mu}^\star$ that would correspond to a fully connected network with uniform weights, $a_{k,\ell}=b_{k,\ell}=1/S$ for all $k,\ell=1,2,\dots,S$.
We start by stating two lemmas (their proofs are given in the sequel).

\vspace*{5pt}
\noindent
\textsc{Lemma 1}
Define an auxiliary function $f_1(t)$ whose values over the negative and positive ranges of time are scaled as follows: 
\beq
f_1(t)=
\frac{t^2}{2} \times 
\left\{
\begin{array}{l}
\displaystyle
{
\max_{\tau\in[0, t]} \left(\frac{d}{d\tau}\frac{\psi(\tau)}{\tau}\right), \quad t\geq 0,
} \\
\\
\displaystyle{
\max_{\tau\in[t, 0]} \left(\frac{d}{d\tau}\frac{\psi(\tau)}{\tau}\right), \quad t<0.
}
\end{array}
\right.
\label{eq:f1def}
\eeq
Then, the LMGF of $\by_{k,\mu}^\star$ for the fully connected solution with uniform weights is:
\beq
\boxed{
\phi_{\mu}^{(c)}(t)=
\frac{S}{\mu}\,
\left[
\int_{0}^{\frac\mu S \,t} \frac{\psi(\tau)}{\tau}d\tau
+
\sum_{i=1}^\infty c_i(t;\mu)
\right]
}
\eeq
where the functions $c_i(t;\mu)$ are nonnegative and satisfy
\beq
\sum_{i=1}^\infty c_i(t;\mu)\leq
f_1\left(\frac\mu S \,t\right)\times\frac{\mu^2}{1-(1-\mu)^2}.
\eeq
~\hfill$\square$

\vspace*{5pt}
\noindent
\textsc{Lemma 2}
Let $\lambda_2$ be the second largest eigenvalue of $A$ in magnitude, and let $|\lambda_2|<\lambda<1$. 
Define another auxiliary function as:
\beq
f_2(t)=
|t| \times 
\left\{
\begin{array}{l}
\displaystyle{\max_{\tau\in[0, t]} |\psi^\prime(\tau)|, \quad t\geq 0,} \\
\displaystyle{\max_{\tau\in[t, 0]} |\psi^\prime(\tau)|, \quad t<0.}
\end{array}
\right.
\label{eq:f2def}
\eeq
Then, the LMGF of the steady-state diffusion output $\by_{k,\mu}^\star$ defined by~(\ref{eq:Theo1}) is:
\beq
\boxed{
\phi_{k,\mu}(t)=
\phi^{(c)}_{\mu}(t)
+
\sum_{i=1}^\infty \sum_{\ell=1}^S  c_{i,\ell}(t;\mu)
}
\eeq
where the functions $c_{i,\ell}(t;\mu)$ now satisfy 
\beq
\sum_{i=1}^\infty 
\sum_{\ell=1}^S |c_{i.\ell}(t;\mu)|\leq
({\cal C}\lambda S)\,\frac{f_2(\mu t)}{1-\lambda(1-\mu)},
\eeq
for a positive constant ${\cal C}$ depending on $\lambda$ and on the combination matrix $A$.
~\hfill$\square$ 

\vspace*{10pt}
\noindent
We can easily show that:
\beq
0\leq f_1(t)<\infty,\qquad 0\leq f_2(t)<\infty,\qquad \forall t\in\mathbb{R}.
\eeq 
Indeed, $f_1(t)\geq 0$ from~(\ref{eq:posder}), while $f_2(t)\geq 0$ by definition. Finiteness of both functions follows from Weierstrass extreme value theorem~\cite{RudinBook} since, by the properties of $\psi(t)$ discussed at the beginning of this appendix, the maxima appearing in~(\ref{eq:f1def}) and~(\ref{eq:f2def}) are maxima of continuous functions over compact sets for any finite $t$.

\vspace*{10pt}
\noindent
By using the above lemmas (whose proofs will be given soon), it is straightforward to prove Theorem 3.

\vspace*{5pt}
\noindent
{\em Proof of Part $i)$ of Theorem~3:} we start by proving that 
\beq
\lim_{\mu\rightarrow 0}\mu\,
\phi^{(c)}_\mu(t/\mu)=S\,\int_0^{t/S} \frac{\psi(\tau)}{\tau}d\tau.
\eeq
From the above Lemma 1 we have:
\beqa
\lefteqn{
\left|
\mu\,
\phi^{(c)}_\mu(t/\mu)-S\,\int_0^{t/S} \frac{\psi(\tau)}{\tau}d\tau
\right|
}
\nonumber\\
&=&
S\,
\,\sum_{i=1}^\infty c_i(t / \mu ;\mu)
\leq
S\,f_1(t/S)\times \frac{\mu^2}{1-(1-\mu)^2}\stackrel{\mu\rightarrow 0}{\longrightarrow 0}.\nonumber\\
\eeqa
On the other hand, using Lemma 2,
\beqa
\lefteqn{\hspace*{-40pt}
\mu\,
\left|
\phi_{k,\mu}(t/\mu)-
\phi^{(c)}_\mu(t/\mu)
\right|
=
\mu\,
\left|
\sum_{i=1}^\infty\sum_{\ell=1}^S c_{i,\ell}(t/\mu;\mu)
\right|
}
\nonumber\\
&\leq&
({\cal C}\lambda S)\,f_2(t)\,
\frac{\mu}{1-\lambda (1-\mu)}\stackrel{\mu\rightarrow 0}{\longrightarrow 0},
\eeqa
and claim $i)$ is proven.

\vspace*{5pt}
\noindent
{\em Proof of Part $ii)$ of Theorem~3:} 
From the definition of $\omega(t)$ in~(\ref{eq:omegadef}) we have $\omega^\prime(t)=\psi(t)/t$, which follows by continuity of $\psi(t)/t$ for all $t\in\mathbb{R}$ --- see property 2) at the beginning of this appendix. 
Then, using the result proven in part $i)$, since $\omega(t)$ is differentiable in $\mathbb{R}$, the G\"artner-Ellis Theorem~\cite{DenHollander} stated in Sec.~\ref{subsec:LDP} can be applied to conclude that $\by_{k,\mu}^\star$ must obey the LDP~(\ref{eq:LDPdef}) with rate function given by the Fenchel-Legendre transform of the function
$S\,\omega(t/S)$.
It is straightforward to verify that the Fenchel-Legendre transform of a function scaled in this way is simply
$S\,\Omega(\gamma)$.

\vspace*{5pt}
We now prove the two lemmas.

\vspace*{5pt}
\noindent
{\em Proof of Lemma 1}. 
For the case of a fully connected network with uniform weights, the finite-horizon variable introduced in~(\ref{eq:regimeterm}) reduces to 
\beq
\by_k^{\star}(n)\;=\;
\sum_{i=1}^n \sum_{\ell=1}^S \mu (1-\mu)^{i-1} \frac{1}{S} \,\bx_{\ell}^{\prime}(i). 
\eeq
Now since the LMGF is additive for sums of independent random variables, the LMGF of $\by_k^{\star}(n)$ defined above, for any fixed time instant $n$, is given by:
\beq
S\,\sum_{i=1}^n \psi\left((1-\mu)^{i-1} \frac \mu S\,t\right).
\label{eq:centLMGF}
\eeq
First we notice that, if we were able to show that this quantity converges as $n$ goes to infinity, the limit will represent the LMGF, $\phi_{\mu}^{(c)}(t)$,  of the steady-state random variable $\by^\star_{k,\mu}$ in the fully connected case, in view of the continuity theorem for the moment generating functions~\cite{CurtissMGF}.
Define $g(t)=\psi(t)/t$ and let us focus initially on $t>0$. We introduce  the countably infinite partition of the interval $[0,\frac \mu S\,t]$ with endpoints
\beq
\tau_i=(1-\mu)^{i-1} \frac \mu S\,t,\qquad i=1,2,\dots,\infty.
\eeq
A second-order Taylor expansion of the function $G(t)=\int_{t}^{\tau_i} g(\tau)d\tau$ around the point $\tau_i$ gives~\cite{RudinBook}: 
\beqa
\int_{\tau_{n+1}}^{\tau_1} g(\tau)d\tau
&=&
\sum_{i=1}^n \int_{\tau_{i+1}}^{\tau_{i}}
g(\tau) d\tau=
\sum_{i=1}^n G(\tau_{i+1})
\nonumber\\
&=&
\sum_{i=1}^n
g( \tau_i)\delta_i
-
\sum_{i=1}^n 
g^\prime(\bar{t}_i) \frac{\delta^2_i}{2}
,\nonumber\\
\eeqa
for a certain $\bar{t}_i\in(\tau_{i+1},\tau_i)$, and with
$
\delta_i=\tau_{i}-\tau_{i+1}.
$
Using the explicit expressions for $\tau_i$ and $g(\cdot)$, we have
\beqa
\sum_{i=1}^n g(\tau_i)\delta_i&=&
\sum_{i=1}^n \psi(\tau_i)\left(1-\frac{\tau_{i+1}}{\tau_i}\right)\nonumber\\
&=&
\mu\,\sum_{i=1}^n \psi\left((1-\mu)^{i-1}\frac \mu S\,t\right),
\eeqa
and we conclude that we can write 
\beq
\mu\,\sum_{i=1}^n \psi\left((1-\mu)^{i-1}\frac \mu S\,t\right)=
\int_{\tau_{n+1}}^{\tau_1} g(\tau)d\tau
+
\sum_{i=1}^n c_{i}(t;\mu),
\label{eq:LMGF2}
\eeq
where $c_i(t;\mu)$ is defined as:
\beq
c_i(t;\mu)=g^{\prime}(\bar{t}_i)\frac{\delta^2_i}{2}>0.
\eeq
Positiveness follows since $g^\prime(t)>0$ for all $t\in\mathbb{R}$ in view of~(\ref{eq:posder}). Now note that 
\beq
\sum_{i=1}^n c_i(t;\mu)
\leq
\sum_{i=1}^\infty \frac{\delta_i^2}{2}\,\max_{\tau \in[0,\mu t/S]} g^\prime(\tau),
\eeq
and recalling the definition of $\delta_i$, we have 
\beqa
\sum_{i=1}^\infty \delta_i^2&=&
\left(\frac \mu S\,t\right)^2
\sum_{i=1}^\infty
[(1-\mu)^{i-1}-(1-\mu)^i]^2\nonumber\\
&=&
\left(\frac \mu S\,t\right)^2\,
\frac{\mu^2}{1-(1-\mu)^2}.
\eeqa
The proof for the case $t<0$ follows the same line of reasoning.
We now obtain
\beq
\sum_{i=1}^\infty c_{i}(t;\mu)\leq f_1\left(\frac \mu S\,t\right) \times \frac{\mu^2}{1-(1-\mu)^2},
\eeq
where $f_1(\cdot)$ is defined in~(\ref{eq:f1def}).
As $n\rightarrow\infty$ in~(\ref{eq:LMGF2}), the first term on the RHS converges to $\int_{0}^{\frac \mu S\,t}g(\tau)d\tau$ since the $\tau_i$'s define a countably infinite partition of $[0,\frac{\mu}{S}\,t]$. The second term is convergent from what was just proved. Using now~(\ref{eq:centLMGF}), and letting $n\rightarrow\infty$, we finally get
\beq
\phi^{(c)}_\mu(t)=\frac{S}{\mu}
\left[
\int_{0}^{\frac \mu S\,t} \frac{\psi(\tau)}{\tau}d\tau
+
\sum_{i=1}^\infty c_{i}(t;\mu)
\right].
\eeq

\vspace*{5pt}
\noindent
{\em Proof of Lemma 2}. Using a first-order Taylor expansion of the function $\psi(\cdot)$, the LMGF of the variable $\by_{k}^{\star}(n)$ defined earlier in~(\ref{eq:regimeterm}) for diffusion networks using combination weights that are not necessarily uniform can be written as:
\beqa
\lefteqn{
\sum_{i=1}^n \sum_{\ell=1}^S \psi\left(\mu(1-\mu)^{i-1}b_{k,\ell}(i)t\right)
}
\nonumber\\
&=&
S\,\sum_{i=1}^n \psi\left((1-\mu)^{i-1}\,\frac \mu S\,t\right)\nonumber\\
&+&
\sum_{i=1}^n
\sum_{\ell=1}^S
\underbrace{\psi^{\prime}(t_{i,\ell})\mu (1-\mu)^{i-1}\,\left[b_{k,\ell}(i)-\frac 1 S\right] t}_{\dfz c_{i,\ell}(t;\mu)},
\nonumber\\
\eeqa
for an intermediate variable $t_{i,\ell}$ that, focusing first on the case $t> 0$, must be certainly contained in the range $[0,\mu t]$, since $b_{k,\ell}\leq 1$. This yields:
\beq
\sum_{i=1}^\infty\sum_{\ell=1}^S |c_{i,\ell}(t;\mu)|\leq
({\cal C}\lambda S)\,\max_{\tau \in[0,\mu t]} |\psi^\prime(\tau)|
 \,
\frac{\mu\,t}{1-\lambda(1-\mu)},
\eeq
where we used Perron's Theorem~(\ref{eq:Perron}). A similar argument holds for $t<0$.

\section*{Appendix C: Convexity properties of $\omega(t)$ and $\Omega(\gamma)$}

\vspace*{5pt}
\noindent
The following properties hold.

\begin{itemize}
\item[$i)$]
$\omega^{\prime\prime}(t)>0$ for all $t\in\mathbb{R}$, implying that $\omega(t)$ is strictly convex.
\item[$ii)$]
$\Omega(\gamma)$ is strictly convex in the interior of the set:
\beq
{\cal D}_{\Omega}=\{\gamma\in\mathbb{R}:\;\Omega(\gamma)<\infty\}.
\eeq
\item[$iii)$]
$\Omega(\gamma)$ attains its unique minimum at $\gamma=\E\bx$, with
\beq
\Omega(\E\bx)=0.
\eeq
\end{itemize}

\vspace*{10pt}
\noindent
{\em Proof.}

\vspace*{10pt}
\noindent
$i)$ In view of~(\ref{eq:omegadef}) we have $\omega^\prime(t)=\psi(t)/t$.
Positivity of $\omega^{\prime\prime}(t)$ follows now from~(\ref{eq:posder}).

\vspace*{10pt}
\noindent
$ii)$ Consider first the following equation:
\beq
\gamma=\omega^\prime(t).
\label{eq:statpointeq}
\eeq
Since $\omega^\prime(t)$ is strictly increasing, it makes sense to define 
\beq
\lim_{t\rightarrow+\infty} \omega^\prime(t)=\omega_{+},
\qquad
\lim_{t\rightarrow-\infty} \omega^\prime(t)=\omega_{-}.
\eeq
Clearly, if $\omega_{+}=+\infty$ and $\omega_{-}=-\infty$, Eq.~(\ref{eq:statpointeq}) will have a solution $t$ for any $\gamma\in\mathbb{R}$.
Consider the most restrictive situation that $\omega_-$ and $\omega_+$ are both finite, and that $\gamma\notin[\omega_{-},\omega_+]$. The case that only one of them is finite follows then in a straightforward manner. 

Recall that the Fenchel-Legendre transform $\Omega(\gamma)$ of the function $\omega(t)$ is defined as:
\beq
\Omega(\gamma)=\sup_{t\in\mathbb{R}}[\gamma t -\omega(t)],
\label{eq:FL}
\eeq
and let us introduce the function:
\beq
h(t)\dfz\gamma t -\omega(t).
\label{eq:hdef}
\eeq
From the first-order condition for strict convexity~(\ref{eq:firstordcond}) applied to the strictly convex function $\omega(t)$, we can write, for $t\neq 0$, $\omega^\prime(t) t >\omega(t)$, which implies:
\beq
h(t)>[\gamma-\omega^\prime(t)]\,t.
\eeq
If $\gamma>\omega_+$, the term on the RHS diverges to $+\infty$ as $t\rightarrow + \infty$.
Similarly, if $\gamma<\omega_-$, the term on the RHS diverges to $+\infty$ as $t\rightarrow - \infty$.
This yields:
\beq
\sup_{t\in\mathbb{R}} h(t)=\infty,
\eeq
showing, in view of~(\ref{eq:FL}) that the condition $\gamma\notin[\omega_{-},\omega_+]$ implies $\gamma\notin{\cal D}_{\Omega}$.

The proof will be complete if we are able to show that $\Omega(\gamma)<\infty$ and $\Omega(\gamma)$ is strictly convex for $\gamma\in(\omega_-,\omega_+)$. 
Now, since $\omega(t)$ is differentiable and strictly convex in $\mathbb{R}$, we have that, for any $\gamma$, the function $h(t)$ in~(\ref{eq:hdef}) is differentiable and strictly concave in $\mathbb{R}$, with
\beq
h^{\prime}(t)=\gamma-\omega^\prime(t).
\eeq
Moreover, for $\gamma\in(\omega_{-},\omega_+)$ the stationary-point equation
\beq
h^{\prime}(t)=0 \Leftrightarrow \gamma=\omega^{\prime}(t)
\label{eq:statpointeq2}
\eeq
admits a unique (since $\omega^{\prime}(t)$ is strictly increasing) solution $t(\gamma)$. 
The strict concavity of $h(t)$ allows us to determine the supremum in~(\ref{eq:FL}) as follows:  
\beq
\Omega(\gamma)=\gamma\,t(\gamma) - \omega(t(\gamma))<\infty,
\label{eq:omegastarprop}
\eeq
where finiteness of $\Omega(\gamma)$ follows by the fact that $t(\gamma)\in\mathbb{R}$, and by finiteness of $\omega(t)$.
By further noting that $\omega^\prime(t)$ is differentiable and $\omega^{\prime\prime}(t)>0$, the theorem about differentiation of the inverse function~\cite[Ex. 2, p. 114]{RudinBook} allows concluding that the derivative of the function $t(\gamma)$ can be computed as:
\beq
\frac{d}{d\gamma}t(\gamma)=\frac{1}{\omega^{\prime\prime}(t(\gamma))}>0.
\eeq
Then we can write
\beq
\frac{d}{d\gamma}\Omega(\gamma)=t(\gamma)+\gamma\,\frac{d}{d\gamma}t(\gamma)-\underbrace{\omega^\prime(t(\gamma))}_{\gamma}\,\frac{d}{d\gamma}t(\gamma)=t(\gamma),
\eeq
and
\beq
\frac{d^2}{d\gamma^2}\Omega(\gamma)=\frac{d}{d\gamma}t(\gamma)>0,
\eeq
which completes the proof.

\vspace*{10pt}
\noindent
$iii)$ We have
\beq
\Omega(\gamma)=\sup_{t\in\mathbb{R}} [\gamma\,t - \omega(t)]\geq \gamma\,0-\omega(0)=0.
\eeq
Since $\omega^{\prime}(0)=\E\bx$, from~(\ref{eq:omegastarprop}) we conclude that
\beq
\Omega(\E\bx)=(\E\bx)\,0 - \omega(0)=0,
\eeq
and, hence, the minimum allowed value of zero is attained.



\end{document}